\documentclass[fleqn,usenatbib]{mnras}

\usepackage{newtxtext,newtxmath}

\usepackage[T1]{fontenc}

\usepackage{graphicx}	
\usepackage{amsmath}	
\usepackage{amssymb}	
\usepackage{enumitem}
\usepackage{subfigure}
\usepackage{booktabs}
\usepackage{threeparttable}

\title{Forbidden line diagnostics of photoevaporative disc winds}

\author[Ballabio et al.]{\parbox{\textwidth}{
G.\,Ballabio$^{1}$\thanks{E-mail: gb258@leicester.ac.uk}, R.\,D.\,Alexander$^1$ and C.\,J.\,Clarke$^2$}\vspace{0.2cm}\\ 
$^{1}$School of Physics and Astronomy, University of Leicester, Leicester, LE1 7RH, UK\\
$^{2}$Institute of Astronomy, Madingley Road, Cambridge CB3 0HA, UK\\
}

\date{Accepted 2020 June 17. Received 2020 June 15; in original form 2020 May 12}

\pubyear{2019}

\begin{document}
\label{firstpage}
\pagerange{\pageref{firstpage}--\pageref{lastpage}}
\maketitle

\begin{abstract}
Photoevaporation driven by high energy radiation from the central star plays an important role in the evolution of protoplanetary discs. Photoevaporative winds have been unambiguously detected through blue-shifted emission lines, but their detailed properties remain uncertain. Here we present a new empirical approach to make observational predictions of these thermal winds, seeking to fill the gap between theory and observations. We use a self-similar model of an isothermal wind to compute line profiles of several characteristic emission lines (in particular the [Ne$\,${\sc{ii}}] line at 12.81 $\mu$m, and optical forbidden lines such as [O$\,${\sc{i}}] 6300\AA~and [S$\,${\sc{ii}}] 4068/4076\AA), studying how the lines are affected by parameters such as the gas temperature, disc inclinations, and density profile. 
Our model successfully reproduces blue-shifted lines with $\varv_{\rm peak} \lesssim 10$~km~s$^{-1}$, which decrease with increasing disc inclination. The line widths increase with increasing disc inclinations and range from $\Delta \varv \sim 15-30$~km~s$^{-1}$. 
The predicted blue-shifts are mostly sensitive to the gas sound speed (and therefore the temperature). The observed [Ne$\,${\sc{ii}}] line profiles are consistent with a thermal wind and point towards a relatively high sound speed, as expected for EUV photoevaporation. However, the observed [O$\,${\sc{i}}] line profiles require lower temperatures, as expected in X-ray photoevaporation, and show a wider scatter that is difficult to reconcile with a single wind model; it seems likely that these lines trace different components of a multi-phase wind. We also note that the spectral resolution of current observations remains an important limiting factor in these studies, and that higher resolution spectra are required if emission lines are to further our understanding of protoplanetary disc winds.
\end{abstract}




\section{Introduction}\label{sec:intro}
Protoplanetary discs are the sites of planet formation, and these discs play a critical role in the formation and early evolution of planetary systems. Planets form in, and from, evolving discs, but consideration of the mass budget shows that only a small fraction of the disc mass ends up in planets; disc evolution is primarily driven by a combination of accretion and mass-loss \citep[e.g.,][and references therein]{2011ARA&A..49..195A, 2014prpl.conf..475A, 2017RSOS....470114E}. However, despite several decades of study the nature of both the accretion and mass-loss processes remains uncertain.

\vspace{1cm}

Observations using a wide variety of different tracers tell us that protoplanetary disc lifetimes are typically a few Myr \citep[e.g.,][]{2001ApJ...553L.153H, 2010A&A...510A..72F, 2012ApJ...745...23M}. Protoplanetary discs are observed to accrete on to their host stars, and the observed accretion rates can plausibly drive disc evolution on Myr time-scales \citep[e.g.][]{1998ApJ...495..385H}. However, discs are also readily observed to be losing mass through jets and winds \citep[e.g.][]{1995ApJ...452..736H}, and the wind mass-loss rates can be comparable to the disc accretion rates \citep{2018ApJ...868...28F}. The relative efficiency of these competing processes remains unclear \citep{2013ApJ...772...60R, 2016ApJ...831..169S, 2019ApJ...870...76B}, but the late-time ``clearing'' of protoplanetary discs is observed to be rapid \citep{1995ApJ...450..824S, 2005ApJ...631.1134A, 2013MNRAS.428.3327K} and very efficient \citep{2006ApJ...651.1177P, 2009ApJ...703L.137I}, which cannot be explained by disc accretion alone. Disc winds are thought to be the dominant driver of protoplanetary disc evolution at late times \citep[e.g.,][]{2001MNRAS.328..485C,2006MNRAS.369..216A,2006MNRAS.369..229A,2010MNRAS.401.1415O}, and may play a major role throughout the disc lifetime. Determining the properties of disc winds is thus a critical step towards understanding planetary systems.

Observations show that disc winds are characterised by two components: a high velocity ($\gtrsim$100\,km\,s$^{-1}$) component associated with magnetically-launched winds/jets, and a low-velocity ($\sim$10\,km\,s$^{-1}$) component that may be either magnetic or thermal in origin \citep{2010MNRAS.406.1553E, 2013ApJ...772...60R, 2016MNRAS.460.3472E, 2018ApJ...868...28F, 2019ApJ...870...76B}. Here we focus on thermal winds, which in this context are invariably driven by photoevaporation. Disc photoevaporation occurs when high energy radiation heats the disc surface to a temperature much higher than that in the disc midplane. Beyond a certain radius the thermal energy of the heated gas is sufficient for it to escape the star's gravitational potential, and the result is a thermally-launched, pressure-driven flow. The characteristic (``gravitational'') radius for disc photoevaporation is
\begin{equation}
    R_{\rm g} = \frac{GM_{*}}{c_{\rm s}^2} \, ,
\end{equation}
where $M_*$ is the stellar mass and $c_{\rm s}$ is the sound speed of the heated gas \citep{1994ApJ...428..654H}. For a 1M$_{\odot}$ star and $c_{\rm s} = 10$\,km\,s$^{-1}$, we find $R_{\rm g} = 8.9$\,au, and more detailed calculations show that the mass-loss rate typically peaks at $\simeq 0.2R_{\rm g}$ \citep{2003PASA...20..337L, 2004ApJ...607..890F, 2007prpl.conf..555D}. Disc photoevaporation is therefore expected to drive significant mass-loss at radii from a few to several tens of au.

For most discs the dominant source of high-energy irradiation is the central star, and three wavelength regimes can be important: ionizing extreme-UV (EUV; 13.6--100eV); non-ionizing far-UV (FUV; 6--13.6eV); and X-rays (0.1--10keV). FUV heating typically only drives mass-loss from the outer regions of the disc ($\gtrsim 50$--100AU; \citealt{2004ApJ...611..360A, 2009ApJ...690.1539G}), and mass-loss rates in these flows depend very sensitively on the disc's radial extent \citep{2019MNRAS.485.3895H}. EUV heating creates an ionized layer on the disc surface, which is approximately isothermal with $c_s \simeq 10$\,km\,s$^{-1}$, and typically drives mass-loss rates $\sim 10^{-10}$M$_{\odot}$\,yr$^{-1}$ \citep{1994ApJ...428..654H,2004ApJ...607..890F}. X-ray irradiation results in slightly cooler flows ($c_s \simeq 3$--5\,km\,s$^{-1}$) but penetrates to higher densities, so can drive mass-loss at rates up to $\sim 10^{-8}$M$_{\odot}$\,yr$^{-1}$ \citep{2010MNRAS.401.1415O,2019MNRAS.487..691P}. It remains uncertain which of these regimes dominates the mass-loss.

The most common way to study these winds is through velocity signatures in emission lines, and the relatively low gas densities mean that forbidden lines are usually the best tracers. Blue-shifted [Ne$\,${\sc{ii}}] 12.81\,$\mu$m emission has been detected from a number of discs \citep{2009ApJ...702..724P,2011ApJ...736...13P,2012ApJ...747..142S}. Due to the high ionization potential of Ne (21.56eV) and the modest critical density of the 12.81\,$\mu$m line (5.0$\times 10^5$\,cm$^{-3}$), these observations represent unambigious detections of low-velocity, low-density, photoionized winds. The observed [Ne$\,${\sc{ii}}] line profiles and fluxes show excellent agreement with theoretical predictions for photoevaporative winds \citep{2008MNRAS.391L..64A,2010MNRAS.406.1553E}, but this line can originate in either EUV- or X-ray-heated winds \citep{2007ApJ...656..515G,2009ApJ...703.1203H,2014prpl.conf..475A} so on its own it does not constrain the mass-loss rate strongly.

Photoevaporative winds also give rise to numerous optical forbidden lines, notably [O$\,${\sc{i}}] 6300\AA, [S$\,${\sc{ii}}] 4068/4076\AA, and [N$\,${\sc{ii}}] 6548\AA~\citep{1995ApJ...452..736H,2014A&A...569A...5N}.
Lines such as [O$\,${\sc{i}}] are readily detected in blue-shifted emission around young stars \citep[e.g.][]{1995ApJ...452..736H, 2013ApJ...772...60R, 2016ApJ...831..169S, 2019ApJ...870...76B}, and the line profiles typically show the two distinct components: a high velocity component (HVC) blue-shifted by $\sim 100$\,km\,s$^{-1}$, and a low velocity component (LVC) blue-shifted by $\sim 10$\,km\,s$^{-1}$ \citep{1995ApJ...452..736H}. In addition, the LVC exhibit a narrow component ($\sim 10$ km/s) and a broad component ($\sim 40$ km/s) \citep{2013ApJ...772...60R}. The high velocity component is unambiguously associated with magnetically-launched jets, but it has long been hypothesised that the LVC traces a photoevaporative disc wind \citep{2004ApJ...607..890F, 2016ApJ...831..169S}. Since it is a neutral line, and O$\,${\sc{i}} has a very similar ionization potential to H$\,${\sc{i}}, we expect little or no [O$\,${\sc{i}}] emission if the flow is fully ionized, and EUV-driven winds fail to reproduce the observed flux of the [O$\,${\sc{i}}] line \citep{2004ApJ...607..890F}. Partially ionized winds, such as those heated by X-rays \citep[e.g.][]{2010MNRAS.401.1415O, 2019MNRAS.487..691P}, can account for the observed line emission, but the nature of the [O$\,${\sc{i}}] LVC remains uncertain: recent studies have suggested both magnetic \citep{2018ApJ...868...28F, 2019ApJ...870...76B} and thermal \citep{2010MNRAS.406.1553E, 2016MNRAS.460.3472E} origins.

The usual approach to modelling emission line profiles from disc winds is to couple numerical hydrodynamic simulations to radiative transfer calculations \citep{2004ApJ...607..890F,2008MNRAS.391L..64A,2010MNRAS.406.1553E,2016MNRAS.460.3472E}. However, these calculations remain computationally expensive, which limits the range of models that can be explored. Moreover, the complexity of these models results in non-linear (and sometimes degenerate) relationships between the physical properties of the wind and the observed line profiles. Consequently there is still a significant gap between these models and observations, and some recent observational studies have instead preferred to take a purely empirical approach fitting line profiles \citep[e.g.,][]{2019ApJ...870...76B}.

In this paper we present a new approach to modelling disc wind observables, which seeks to bridge this gap between theory and observations. We make use of an analytic disc wind solution, which allows rapid calculation of forbidden emission line diagnostics for large numbers of wind models. We then explore in detail how different emission line diagnostics trace the properties of the underlying disc wind, in order to understand what we can - and cannot - expect to determine from current observations. We describe our model in Section \ref{sec:methods}, and present the results of our calculations in Section \ref{sec:results}. In Section \ref{sec:discussion} we discuss the implications of these results, as well as the caveats and limitations of our approach, and we summarize our conclusions in Section \ref{sec:concs}.

\section{Methods}
\label{sec:methods}

\subsection{The Self-similar Model}\label{sec:2dflow}
Most current models of photoevaporative winds use numerical hydrodynamic or radiation-hydrodynamic simulations to calculate the flow structure \citep[e.g.,][]{2004ApJ...607..890F, 2010MNRAS.401.1415O, 2017ApJ...847...11W}. The main reason for this is that no fully self-consistent analytic solutions of the flow structure exist, even in the simplest cases \citep[e.g.,][]{1983ApJ...271...70B}. Recently \citet[][hereafter CA16]{2016MNRAS.460.3044C} derived an analytic model for an axisymmetric, thermally driven disc wind. CA16 studied the topology of an isothermal flow launched from a disc where the density at the flow base is a power law of the radius, and derived self-similar solutions for the 2D flow structure. Their solution is strictly valid only in the limit where gravity and centrifugal forces are negligible, which should limit its applicability to large radii ($R\gg R_{\mathrm g}$). However, CA16 showed that the self-similar solution actually shows good agreement with hydrodynamic simulations even for radii smaller than the gravitational radius of the flow. This is due to the near cancellation of gravitational and centrifugal terms close to the wind base, and because each of these terms falls off more steeply with distance along the streamlines than the dominant terms associated with the convective derivative and with pressure gradients perpendicular to the streamlines. 

Here we compute wind solutions following the approach of CA16; for reference we summarize their solution here. These 2D self-similar solutions are scale-free, and adjacent streamlines in the flow are therefore simply re-scaled versions of one another.  We define $R_{\rm b}$ as the (cylindrical) radius at the base of a streamline, $\rho_{\rm b}$ as the density at this point, and $u_{\rm b}$ as the velocity (in the $R$-$z$ plane) at the streamline base. All the spatial variables are then scaled as multiples of $R_{\rm b}$; the density as a multiple of $\rho_{\rm b}$; and the velocity as a multiple of $u_{\rm b}$. We consider power-law profiles of the base density
\begin{equation}
\rho_0(R) = \rho_{\rm b} \left(\frac{R}{R_{\rm b}}\right)^{-b}
\end{equation}
and compute streamline solutions for an axisymmetric disc wind in a Cartesian coordinate system (x,y). 
We convert the continuity equations for mass and angular momentum into dimensionless (scale-free) form, and (following CA16) use\,\,$\tilde{}$\,\,symbols to denote dimensionless variables. In this form, the scaled conservation of mass flux along a streamline becomes
\begin{equation}\label{eq:massflux}
    \tilde{\rho} \, \tilde{u} \, \tilde{A} = 1,
\end{equation}
where $\tilde{A}$ is the normalized area of a streamline bundle. Similarly, Bernoulli's equation becomes
\begin{equation}
    \tilde{\rho} \, {\rm exp} \left[\frac{u_b^2}{2c_{\rm s}^2}(\tilde{u}^2-1) \right] = 1,
\end{equation}
and the normalized Euler equation perpendicular to the streamlines can be written as 
\begin{equation}
    \frac{\tilde{u}^2 u_{\rm b}^2}{\tilde{R}_{\rm eff}} = c_{\rm s}^2 \nabla \rm{ln} \rho,
\end{equation}
where ${\tilde{R}_{\rm eff}}$ is the local radius of curvature of the streamline (we refer the reader to CA16 for a more complete description).

\begin{figure}
	\centering
	\includegraphics[width=0.45\textwidth]{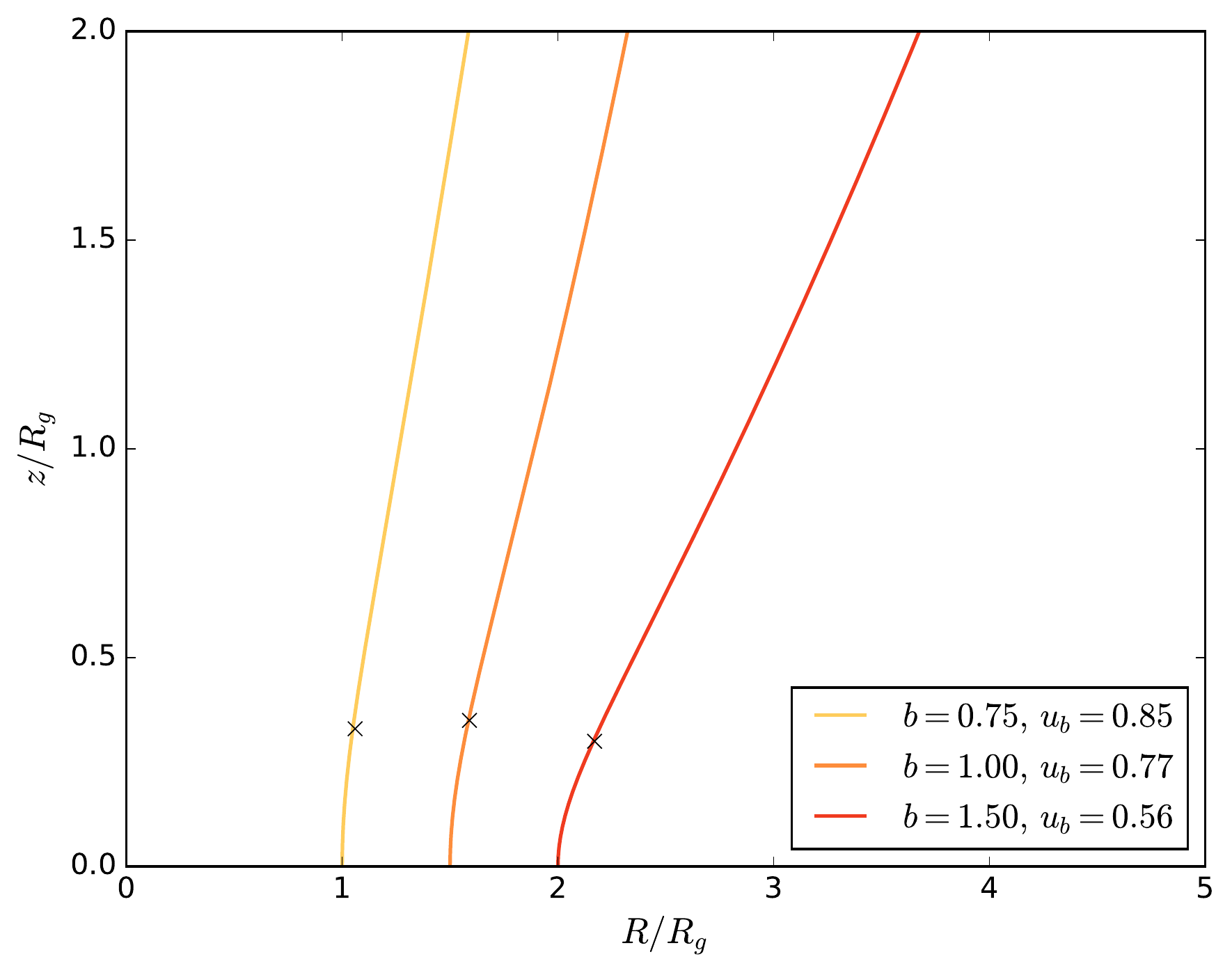}
	\caption{Sample solutions of streamlines derived from the model by CA16. We plot the spatial coordinates ($\tilde{x},\, \tilde{y}$) of each set of streamlines in a polar plane (we introduce an offset along the radial coordinate for clarity). Each streamline topology differs according to the value of $b$ and the corresponding value of $u_{\rm b}$, taken from CA16. The parameter $b$ represents the steepness of the radial density profile, while $u_{\rm b}$ is the launching velocity at the base of the wind. The black cross traces the sonic point for each line.}
	\label{topologystreamlines}
\end{figure}

Within this framework, we construct the streamline topology following the numerical method adopted in CA16. We fix the first point P$_1$ at the base, which in dimensionless units takes the values $\tilde{x}_1=1, \, \tilde{y}_1=0, \, \tilde{x}'_1=0$ and $\tilde{u}_1=u_{\rm b}$ \footnote{Here the subscripts 1 and 2 refer to variables evaluated at P$_1$ and P$_2$, respectively}. For these known values, we then use Eq.~(16) in CA16 to determine $\tilde{u}'_1$, and build the streamline structure by integrating along the streamline. We take $\tilde{y}$ as the independent variable and use a Runge-Kutta method, with step size $\Delta \tilde{y}$, to calculate $\tilde{u}_2$ at the subsequent point P$_2$ along the streamline.
We then use Eq.~(14) and Eq.~(15) in CA16 to calculate $\tilde{x}''_1$ at P$_1$, and hence determine the spatial coordinates and the local streamline gradient at P$_2$ as:
\begin{align}
& \tilde{x}_2= \tilde{x}_1 + \tilde{x}'_1 \Delta \tilde{y} + 0.5 \tilde{x}''_1 \Delta \tilde{y}^2 ,\\
& \tilde{x}'_2= \tilde{x}'_1 + \tilde{x}''_1 \Delta \tilde{y} .
\end{align}
and then compute the density at each point on the streamline using Eq.~\ref{eq:massflux}.
The resulting solutions have just two free parameters: the power-law index $b$, and the launch velocity $u_{\rm b}$. However, for a given $b$ there is a maximum allowed value of $u_{\rm b}$, and CA16 also showed that this maximum value provides the best agreement with numerical solutions. We therefore follow CA16 and adopt the maximum value of $u_{\rm b}$ in all our calculations, and we refer the reader to CA16 for a complete derivation of the analytic solution.
We also note that no self-similar solutions exist for very steep density profiles ($b\geq2$)\footnote{For very steep density profiles the base density declines more rapidly with radius than the density in the flow. This results in the streamlines becoming convex upwards, which violates the self-similar assumption.}.
With this approach the solutions are therefore completely described by the power-law index $b$. We plot the streamlines for different values of $b$ in Fig.~\ref{topologystreamlines}, within the range 0.75-1.5. This is similar to the range of density profiles in the line-emitting region in previous models \citep[e.g.,][]{2009ApJ...703.1203H,2010MNRAS.406.1553E} and, as we expect the forbidden line emission to scale approximately $\propto n^2$, is also roughly consistent with the range of radial spectral indices derived by \cite{2016ApJ...831..169S}.

\subsection{Generating Line Profiles}
\label{sec:lineprof}
Our aim in this paper is to use these self-similar solutions to compute empirical diagnostics of photoevaporative winds, and here we focus on the forbidden emission lines that have previously been shown to be good tracers of photoevaporation \citep[e.g.,][]{2004ApJ...607..890F, 2008MNRAS.391L..64A, 2010MNRAS.406.1553E}. We use the 2D flow solutions described in Section \ref{sec:2dflow} to compute the global wind structure, and then compute the line profiles. For a chosen parameters $b$ and $u_{\rm b}$, the self-similar solution provides the density and velocity at each point along the streamline. We compute these solutions on a 2D polar grid, logarithmically spaced along the radial direction and linearly spaced along the angular direction, such that the grid cells are approximately square (i.e. $\Delta r=r\Delta\theta$). We use $N_r =1113$ cells along the radial direction and $N_{\theta} = 250$ along $\theta$, spanning a range $r = [0.03, 10] \, R_{\rm g}$ and $\theta = [\pi/6, \pi/2]$ \footnote{Conventionally, $\theta=0$ is the polar axis, whereas $\theta=\pi/2$ refers to the midplane.}. Since the problem is scale free, we can re-scale the original streamline to a reference value, and assign a density, radial velocity, and polar velocity to each cell of the grid. For a given value of the power-law index $b$ (i.e. for one of the lines in Fig.~\ref{topologystreamlines}) the streamline is monotonic in the polar angle $\theta$, so for each cell on the grid we calculate the polar angle at the centre of the cell. This corresponds to a unique point on the streamline, and allowing us to re-scale the flow variables of the streamline accordingly. 

In order to extend the solutions to 3D, we assume reflective symmetry about the disc mid-plane, and rotational symmetry around the (vertical) $z$-axis. We expand the coordinate range to $\theta = [\pi/6, 5\pi/6]$ and $\phi = [0, 2\pi]$. The resulting 3D grid has $1113 \times 500 \times 1200$ cells in the $r, \theta$ and $\phi$ directions respectively. Tests indicate that our results are not sensitive to the exact numerical resolution. We assume that the wind base is in Keplerian rotation, and set the azimuthal velocity in each cell such as the Keplerian angular momentum is conserved along the streamline. Once a value of density and radial velocity has been assigned to each cell, we calculate the mass loss rate as the mass flow through a surface at a fixed radius. Integrating over the polar angle yields an estimate of the mass loss rate of a wind flowing from a certain region. As the analytic solutions are scale-free, we express the density normalisation in terms of the equivalent total mass loss rate within a radius of 25 au. The value of the mass loss rate is one of the key parameters of the model, as will be shown later.

From these density and velocity fields we then compute the properties of forbidden lines, by assuming the emission is optically thin and following the approach used in \citet{2004ApJ...607..890F}, \citet{2007ApJ...656..515G} and  \citet{2008MNRAS.391L..64A}. The observed line luminosity as a function of velocity $v$ is given by
\begin{align}\label{eq:lineprofile}
    L(v) \, = \, &\frac{1}{\sqrt{2\pi}v_{\rm th}} \int_{\rm V} {\rm exp} \left\{ -\frac{[v - v_{\rm los}(\textbf{r})]^2}{2v_{\rm th}^2} \right\} Ab_{\rm X} \, X_{\rm j} \, n_{\rm e}(\textbf{r}) \nonumber \\
    &\times \, P_{\rm u} \, A_{\rm ul} \, h \nu_{\rm ul} \, \rm dV, 
\end{align}
where the integral is evaluated over each cell in the grid. Here $n_{\rm e}(\textbf{r})$ is the number density of the gas and $h \nu_{\rm ul}$ is the energy of emitted photons. $A_{\rm ul}$ is the Einstein coefficient for the transition. We adopt the standard solar values for the abundances $Ab_{X}$ \citep{2011piim.book.....D}. $X_{\rm j}$ denotes the fraction of species $X$ that exists in the $j$th ionization state. As our wind solutions are isothermal and optically thin, we treat $X_{\rm j}$ as a constant.  The values of $Ab_{X}$ and $X_{\rm j}$ therefore affect only the absolute line luminosity, and have no effect on the (normalized) line profiles. Values for the lines we consider are given in Table~\ref{tab:atomicparams}. 
$v_{\rm los}$ is the gas velocity component along the line of sight, calculated for disc inclination angle $i$ as
\begin{align}
    v_{\rm los} = \, & [{\rm cos}\theta \, {\rm cos}\phi \, {\rm sin} \, i + {\rm sin}\theta \, {\rm cos} \, i] \, v_r \nonumber \\
                  & [-{\rm sin}\theta \, {\rm cos}\phi \, {\rm sin} \, i + {\rm cos}\theta \, {\rm cos} \, i ] \, v_{\theta} \nonumber \\
                  & [-{\rm sin}\phi \, {\rm sin} \, i] \, v_{\phi} \, .
\end{align}
The inclination angle is defined as the angle between the line of sight and the perpendicular to the disc, i.e. $i=\pi/2$ corresponds to an edge-on disc, while $i=0$ corresponds to a face-on disc. The thermal velocity $v_{\rm th}$ for species $X$ is given by
\begin{equation}
    v_{\rm th} = c_{\rm s} \sqrt{m_{\rm H} / m_{\rm X}}.
\end{equation}
The sound speed is strictly related to the gas temperature and it is also one of the crucial parameters of our model, as we will see.
The excitation fraction $P_{\rm u}$ is evaluated as
\begin{equation}
    P_{\rm u} = \frac{1}{2 C_{\rm ul} {\rm exp} (T_{\rm ul}/T) +1},
\end{equation}
where $T_{\rm ul}$ is the excitation temperature for each transition and $T$ is the gas temperature. The term $C_{\rm ul}$ describes the departure from a thermal level population, and can be expressed as 
\begin{equation}
    C_{\rm ul} = 1 + \frac{n_{\rm cr}}{n_{\rm e}(\textbf{r})},
\end{equation}
where $n_{\rm cr}$ is the critical density for the transition \citep{2007ApJ...656..515G}.
\begin{table}
    \centering
    \caption{Atomic constants for the line transitions considered in this study. The [Ne$\,${\sc{ii}}] atomic constants are taken from \citet{2007ApJ...656..515G}. The other atomic constants are taken from \citet{2014A&A...569A...5N}. The parameters are ordered by the higher critical density.}
    \begin{tabular}{c|c|c|c|c|c}
    \toprule
         Ions & $\lambda$ & $Ab_{\rm X}$\textsuperscript{*} & $A_{\rm ul}$ & $T_{\rm ul}$ & $n_{\rm cr} $ \\
          & [\AA] &  & $[\rm{s}^{-1}]$ & $[\rm{K}]$ & $[\rm{cm}^{-3}]$ \\
    \midrule
         S$\,${\sc{ii}} & 4068.6 & 1.45e-5 & 1.9e-1 & 35354 & 2.6e6 \\
         S$\,${\sc{ii}} & 4076.3 & 1.45e-5 & 7.7e-2 & 35430 & 1.9e6 \\
         O$\,${\sc{i}} & 6300.3 & 5.37e-4 & 5.6e-3 & 22830 & 1.8e6 \\
         Ne$\,${\sc{ii}} & 128100 & 1.0e-4 & 8.39e-3 & 1122.8 & 5.0e5 \\
         S$\,${\sc{ii}} & 6716.4 & 1.45e-5 & 2.0e-4 & 21416 & 1.7e3 \\
    \bottomrule \\
    \multicolumn{6}{l}{\textsuperscript{*}\footnotesize{Solar abundances taken from \cite{2011piim.book.....D}, following}} \\
    \multicolumn{6}{l}{\footnotesize{\, \cite{2009ARA&A..47..481A}.}} \\
    \multicolumn{6}{l}{} \\
    \end{tabular}
    \label{tab:atomicparams}
\end{table}

\subsection{Comparison with hydrodynamic simulations}
\label{sec:hydro_comp}
\begin{figure*}
	\centering
	\includegraphics[width=0.8\textwidth,trim={0cm 0cm 0cm 0cm},clip]{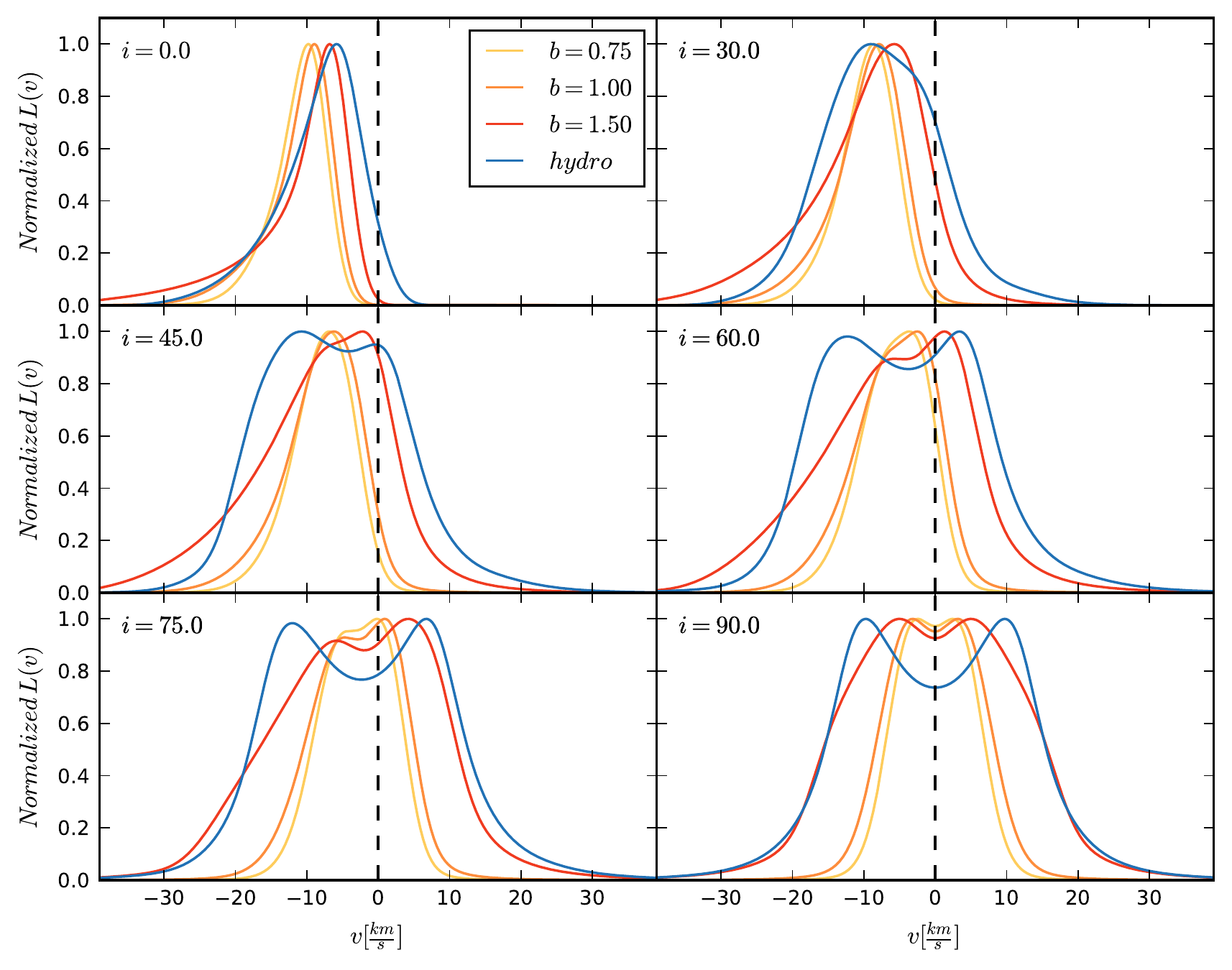}
	\caption{Theoretical line profiles for the [Ne$\,${\sc{ii}}] emission line at $12.81 \, \mu \rm{m}$ at different disc inclinations. The yellow, orange and red lines are the line profiles obtained from the wind streamlines shown in Fig.~\ref{topologystreamlines}, for $b=$0.75, 1.00 and 1.50. The line profiles are computed for a 10\,km\,s$^{-1}$ disc wind model launched from a region between 0.1 and 9.5 $\rm{R}_{\rm g}$ along the radial direction and up to $z\sim 10\,\rm{R}_{\rm g}$ in the vertical direction. The emission is calculated with density normalized to $n_g$ at $R=\rm{R}_{\rm g}$, as in the hydrodynamic simulations.
	For comparison, in blue we show the hydrodynamic solutions taken from \citet{2011ApJ...736...13P}, where they consider a $1\rm{M}_{\odot}$ star with ionizing luminosity $\Phi=10^{41}$ photons s$^{-1}$ and $n_{\rm g}=\rho_{\rm g}/\mu m_{\rm H} = 2.8 \times 10^4 \rm{cm}^{-3}$. Zero km/s is represented by the vertical black dashed line.}
	\label{line_profile_comp}
\end{figure*}
To test our method, and the suitability of the self-similar solutions for line profile calculations, we first perform a comparison with existing hydrodynamic simulations. We compute profiles of the well-studied [Ne$\,${\sc{ii}}] fine-structure line at 12.81\,$\mu \rm{m}$ (see Table~\ref{tab:atomicparams} for the atomic parameters of the transition), using the method described in Section~\ref{sec:lineprof}. For comparison we also compute line profiles using the hydrodynamic simulation from \citet{2011ApJ...736...13P} as input, instead of the self-similar solution. The base density profile in this simulation follows the broken power-law form of \citet{2004ApJ...607..890F} and \citet{2008MNRAS.391L..64A}, based on the radiative transfer calculations of \citet{1994ApJ...428..654H}. This prescription assumes a density profile that transitions smoothly from $b=1.5$ at $R<R_{\rm g}$ to $b=2.5$ at $R>R_{\rm g}$ (though in practice most of the [Ne$\,${\sc{ii}}] line flux comes from $R < R_{\rm g}$, where $b=1.5$). 

Figure~\ref{line_profile_comp} shows the line profiles at different disc inclinations, with the yellow, orange and red lines representing the line emission derived from the analytical model for $b=$ 0.75, 1.0 and 1.5 respectively, of a wind launched from a disc with temperature $T\sim 10^4$~K ($c_{\mathrm s}=10$~km\,s$^{-1}$). The streamlines are sampled over a region between 0.1 and 9.5\,$\rm{R}_{\rm g}$ along the radial direction and up to $z \sim$\,10\,$\rm{R}_{\rm g}$ in the vertical direction. The blue line shows the results from the hydrodynamic calculation, for a $1\rm{M}_{\odot}$ star with ionizing luminosity $\Phi=10^{41}$ photons s$^{-1}$. For this comparison (only) we normalise all the self-similar solutions to have the same density at $R=\rm{R}_{\rm g}$.

We anticipate minor differences between the two methods, as both the base density profile and streamline topology differ, and we do indeed see such differences in the line profiles in Fig.\,\ref{line_profile_comp}. The power-law density profiles from the analytic model result in significantly higher density in the wind at large radii ($R\gg R_{\mathrm g}$), and the total mass-loss rate actually diverges to large $R$ for $b<2$. Indeed, by considering our model out to $\sim$\,10\,$R_{\rm g}$ rather than a broken power law, we are over-weighting the emission from the outer disc and, as a consequence, there is a higher contribution to the line centre at high disc inclinations. The profiles therefore vary somewhat with our choice of outer radius, and for high inclination angles we find that larger outer radii result in slightly narrower lines, as there is more emission at low rotational velocity. For lower inclinations we also see a slightly more emission at higher blue-shifts as the outer radius is increased since, as we compute the lines from a spherical grid, a larger outer radius results in the inclusion of material at high $z$ where the flow velocity is somewhat higher. These are not strong effects, however, and large outer radii ($\gtrsim$10$R_{\mathrm g}$) are in any case disfavoured, as they correspond to unrealistically large launching radii ($\gtrsim 100$\,au). 

At small radii, $R < R_{\mathrm g}$, the gas density near the base of the flow often exceeds the critical density. This reduces the flux emitted from this region. If, however, the density was sub-critical everywhere, the line profiles would essentially be scale-free, but $n > n_{\mathrm {cr}}$ at small radii in all our calculations, so there are always some departures from self-similarity.  This primarily affects the lines at low inclinations, where the relative suppression of emission at super-critical densities leads to (relatively) more flux on the blue wing of the line, and is more pronounced for steeper base density profiles (i.e., larger values of $b$), which have higher densities at small $R$.  

The combination of the effects discussed above means that our line profiles vary slightly with both the density normalisation and the choice of outer radius, and the self-similar solutions generally show narrower lines at high inclinations than those from the hydrodynamic calculations. These differences are most apparent in the line wings (particularly when we consider normalised line profiles), but they are small effects. A secondary effect is that the hydrodynamic calculations launch off an inclined plane, while the theoretical wind models launch off the disc midplane, and this reduces the blue-shift of the line peak by $\sim 1\, \rm{km \, s^{-1}}$. 

Overall the [Ne\,{\sc ii}] line profiles from the analytic model with index $b=1.50$ show good agreement with those from the hydrodynamic simulations, particularly when we consider the most readily observable parameters such as the line-width and peak velocity. The main differences are seen at high inclination angles, where (as noted above) the larger outer radius in the analytic calculations leads to a less pronounced double-peak. Moreover, while these small differences are notable in theoretical calculations (with unlimited spectral resolution) they are unlikely to be measurable in real observations, so we conclude that our analytic model provides a reliable method for studying the observational properties of emission lines from thermal disc winds.

\subsection{Parameters}
The principal advantage of the analytic wind solutions is that they allow efficient exploration of a large physical parameter space without requiring computationally expensive hydrodynamic calculations. Our solutions have 6 free parameters: the power-law index of the base density profile, $b$; the inner ($R_{\rm in}$) and and outer ($R_{\rm out}$) radii of the flow base; the sound speed $c_{\rm s}$; the normalization of the density; and the disc inclination $i$. 
These parameters, introduced in Section~\ref{sec:2dflow}, determine the physical properties of the thermal wind.
$b$ specifies the density profile at the base of the wind, regulating also the wind density structure and its emission. A steeper density profile (higher value of $b$) has higher density at smaller radii, resulting in relatively more emission coming from the inner regions. The index $b$ also sets the launch velocity $u_{\mathrm b}$ (as noted in Section~\ref{sec:2dflow}): lower values of $b$ result in higher wind velocities, and correspondingly larger peak velocities in the line profiles (see Fig.\ref{line_profile_comp}).  As we choose to normalise the density to the mass-loss rate, rather than an absolute density value, the value of $\dot{M}$ also affects the density distribution and the resulting emission. Forbidden lines predominantly originate in those regions where the gas density is smaller than the critical density. Increasing the mass-loss rate moves the critical density contour outwards, and suppress the line emission from smaller radii. 
For the lines considered here, the gas density at the base of the innermost streamlines is always super-critical. Our results are therefore not very sensitive to the choice of inner radius, and extending the models to $R_{\mathrm {in}} < 0.1R_{\mathrm g}$ has a negligible effect on the line profiles.
Finally, the sound speed $c_{\mathrm s}$ of a photoevaporative wind is determined largely by the irradiating spectrum. EUV radiation typically heats the disc at a temperature of $T \sim 10^4$ K, which corresponds to a sound speed of 10~km\,s$^{-1}$. X-ray-heating instead heats the gas to $T \sim 2000-5000$ K, which corresponds to $c_{\mathrm s}=3-5$ km\,s$^{-1}$.

We explore a range of values for these parameters, which are listed in Table~\ref{tab:modelparams}.
\begin{table}
    \centering
    \caption{Range of parameters explored in this study. $b$ represents the steepness of the radial density profile at the base of the wind. The wind is launched from a region with inner and outer boundaries [$R_{\rm in}$, $R_{\rm out}$] and a gas sound speed $c_{\mathrm s}$. We express the density normalisation in terms of the equivalent mass loss rate within a radius of 25 au.}
    \begin{tabular}{c|c|c|c|c}
    \toprule
         $b$ & $R_{\rm in}$ [R$_{\rm g}$] & $R_{\rm out}$ [R$_{\rm g}$] & $c_{\rm s}$ [$\rm{km \, s^{-1}}$] & $\dot{M}$ [${\rm M}_{\odot}$/yr] \\
    \midrule
         0.75 & 0.01 & 5.0 & 3.0 & 10e-10 \\
         1.00 & 0.03 & 10.0 & 5.0 & 10e-9 \\
         1.50 & 0.1 &  & 10.0 & 10e-8 \\
    \bottomrule \\
    \end{tabular}
    \label{tab:modelparams}
\end{table}
For each set of parameters we compute the emission for disc inclinations spanning the range $[0, \pi/2]$. We also compute the line profiles for different species, starting with the [Ne$\,${\sc{ii}}] forbidden line at $12.81 \, \mu \rm{m}$ (discussed in Section \ref{sec:hydro_comp}). We also consider the optical forbidden lines of [S$\,${\sc{ii}}] at 4068/4076\AA~and , and the (neutral) [O$\,${\sc{i}}] 6300\AA~line. Optical forbidden lines have also been used extensively as wind diagnostics \citep[e.g.][]{1995ApJ...452..736H,2014A&A...569A...5N}; we focus on these lines in particular as they are representative tracers which are sensitive to different regions of the flow. The high critical density and low excitation temperature of the [Ne$\,${\sc{ii}}] line makes it an excellent tracer of the launching region of photoevaporative winds \citep[e.g.][]{2008MNRAS.391L..64A, 2010MNRAS.406.1553E}, while the [S$\,${\sc{ii}}] 4068/4076\AA~doublet has a similar critical density but is more sensitive to the gas temperature. [S$\,${\sc{ii}}] 6716\AA~has a much lower critical density, while the well-studied [O$\,${\sc{i}}] 6300\AA~line arises in neutral gas and is therefore a key diagnostic of the ionization state. The atomic parameters for these lines are listed in Table~\ref{tab:atomicparams}. 

\hspace{2cm}

\section{Results}
\label{sec:results}
The method presented here successfully allows us to compute the line profiles of different wind tracers. It is much simpler than the previously adopted methods \citep{2010MNRAS.406.1553E, 2016MNRAS.460.3472E}, which require the use of computationally expensive hydrodynamic and/or radiative transfer simulations. Despite the significant simplifications of our model, we generally find good agreement with both more sophisticated models \citep[e.g.][]{2004ApJ...607..890F, 2008MNRAS.391L..64A, 2010MNRAS.406.1553E, 2016MNRAS.460.3472E}, and with observations \citep{2009ApJ...702..724P, 2012ApJ...747..142S}, with blue-shifts of a few to $\simeq$10\,km\,s$^{-1}$ which vary strongly with disc inclinations, as shown in Fig.~\ref{line_profile_comp}. Since most of the emission comes from the inner regions and the velocity there is subsonic, we expect to measure line peaks of the order of the sound speed.
The line profiles presented in Fig.~\ref{line_profile_comp} are at `infinite' resolution, so in order to consider their observable properties we degrade our line profiles to the typical resolution of an echelle spectrograph. To do this we convolve the line profiles calculated from the model with a Gaussian profile of half width $\sigma = 5 \, \rm{km \, s^{-1}}$, which corresponds to a spectral resolution R$=\lambda/\Delta \lambda=30\,000$ (approximately the resolution of current echelle spectrographs, such as VISIR on the ESO Very Large Telescope; e.g., \citealt{2012ApJ...747..142S}). 

\begin{figure}
	\centering
	\includegraphics[width=0.47\textwidth]{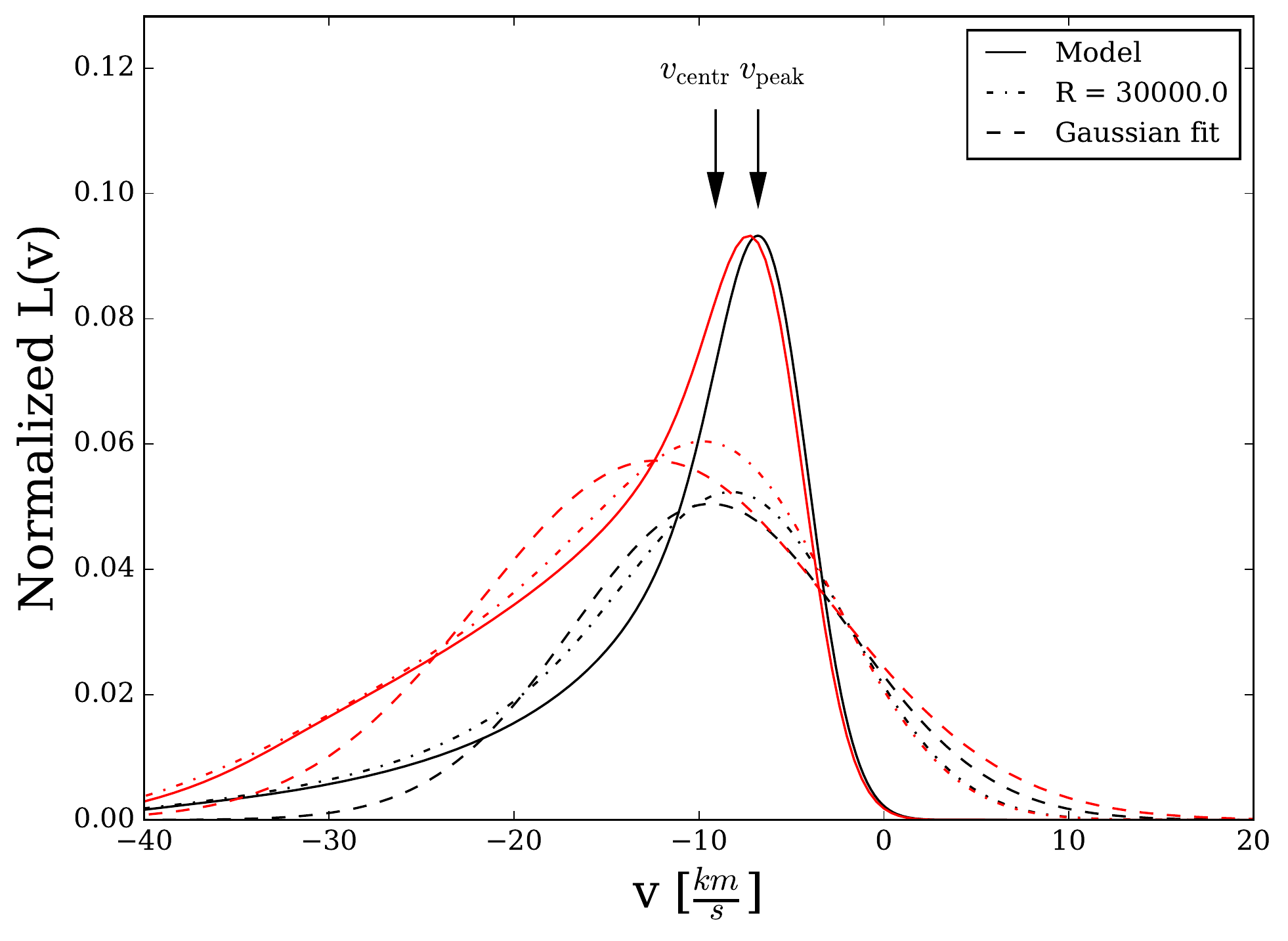}
    \caption{Theoretical [Ne\,{\sc ii}] line profile for the $b=1.50$ face-on disc shown in Fig.~\ref{line_profile_comp} (solid line). The line is degraded to a spectral resolution R$=30\,000$ (dash-dotted line) and then fitted to a single Gaussian profile (dashed line). We observe how changing the emitting domain affects the line shape. The black and red lines show the emission line profiles from a region extended up to z~$\sim10$~R$_{\rm g}$ and z~$\sim100$~R$_{\rm g}$, respectively. The red line profile is normalised such that it has the same peak of the black solid line profile - all the red profiles have been normalised accordingly. The arrows show the position of the peak and centroid velocities of the theoretical line. The properties of the line profiles are listed in Table~\ref{tab:valueslines}.}
    \label{fig:line_convolfit}
\end{figure}

Our main aim is to provide insight into current and future studies of emission lines, so we focus on deriving common observables, particularly the blue-shift and line width. To do this we follow common observational procedure and fit a single Gaussian profile to our model lines, from which we determine the velocity at the peak, and the full width at half maximum flux (FWHM). However, the peak velocity is not a robust diagnostic (especially at high disc inclinations, where the lines are typically double-peaked), so we also compute the velocity of the flux centroid, which is a more reliable estimate of the bulk velocity. As an example, such properties of the line profile are listed in Table~\ref{tab:valueslines}, for the corresponding lines shown in Fig.~\ref{fig:line_convolfit}. The latter illustrates a summary of the observational approach described above, for the line profile from a face-on disc shown in Fig.~\ref{line_profile_comp}. The black line is computed for a 10\,km\,s$^{-1}$ disc wind, with a density power law given by $b=1.50$ and launched from a region between 0.1 and 9.5 $\rm{R}_{\rm g}$ along the radial direction and up to $z\sim 10\,\rm{R}_{\rm g}$ in the vertical direction. The solid line indicates the theoretical line profile, the dash-dotted line is the convolution at spectral resolution R$=30\,000$, and the dashed line is the Gaussian fit to the convolved line profile. The black arrows highlight the difference between the peak and centroid velocities. 
\begin{table}
    \centering
    \caption{Properties of the line profiles shown in Fig.~\ref{fig:line_convolfit}. In black and red are the velocity at peak, centroid velocity and FWHM corresponding to the emission line profiles from a region extended up to z~$\sim10$~R$_{\rm g}$ and z~$\sim100$~R$_{\rm g}$, respectively.}
    \begin{tabular}{l|cc|cc|cc}
    \toprule
        & \multicolumn{2}{c}{$v_{\rm peak}$ [$\rm{km \, s^{-1}}$]} & \multicolumn{2}{c}{$v_{\rm centr}$ [$\rm{km \, s^{-1}}$]} & \multicolumn{2}{c}{FWHM [$\rm{km \, s^{-1}}$]} \\
    \midrule
        Model & -6.8 & \textcolor{red}{-7.2} & -9.1 & \textcolor{red}{-11.9} & 7.6 & \textcolor{red}{11.2} \\
        Convolution & -8.3 & \textcolor{red}{-9.6} & -10.1 & \textcolor{red}{-12.8} & 16.4 & \textcolor{red}{20.8} \\
        Gaussian fit & -9.4 & \textcolor{red}{-12.4} & -9.4\textsuperscript{*} & \textcolor{red}{-12.4}\textsuperscript{*} & 16.9 & \textcolor{red}{18.9} \\
    \bottomrule \\
    \multicolumn{7}{l}{\textsuperscript{*}\footnotesize{The centroid velocity, by definition, is the same as the peak velocity}} \\
    \multicolumn{7}{l}{\footnotesize{\, for the fitted lines.}} \\
    \multicolumn{7}{l}{} \\
    \end{tabular}
    \label{tab:valueslines}
\end{table}

It is notable that the observational procedure leads to a non-negligible shift in the measured properties of the lines, mainly due to the  spectral resolution. Even in the case of a face-on disc, where the line is single-peaked, the profile is markedly asymmetric, with an extended wing on the blue side. At $R=30,000$ the instrumental broadening is such that a (symmetric) Gaussian is still a relatively good fit, but at higher resolution this asymmetry could introduce a greater uncertainty to the fitting. 
The marked blue wing, and its effects on the observational procedure, are enhanced when we consider the emission from a wider region. The self-similar solutions can be extended to arbitrarily large radii, and therefore we consider the results varying the spherical radius of the domain. In Fig.~\ref{fig:line_convolfit} we compare the line profiles of a wind launched from a region extending up to z~$\sim10$~R$_{\rm g}$ (in black) and to z~$\sim100$~R$_{\rm g}$ (in red). In order to highlight the discrepancies in the two profiles, the red line profile is normalised such that the two theoretical line profiles have the same peak - all the red profiles have been normalised accordingly. 
The values of peak velocity, centroid velocity, and FWHM for each corresponding line profile are given in Table~\ref{tab:valueslines}. Although the peak velocity is not strongly affected, the line wing ``pulls'' the centroid velocity, and the fitted peak, to larger blue-shifts as the size of the emitting region is increased. The line FWHM also increases, due to the higher wind velocities at large $z$. However, these shifts are relatively small, and this is an extreme case: $100$~R$_{\rm g}\simeq900$~au for a 1M$_{\odot}$ star, and in real systems we do not expect to see a significant contribution to the emission from such large radii.  Moreover, Table~\ref{tab:valueslines} shows that uncertainty in the peak/centroid velocity due to fitting a Gaussian profile to the asymmetric lines is as large as the uncertainty due to the size of the emitting region.  We therefore consider only the smaller domain (up to z~$\sim10$~R$_{\rm g}$) for our subsequent calculations.

Furthermore, we use a chi-squared test to determine whether there is a significant difference between the convolved line profile and the Gaussian fit. We then calculate the minimum chi-squared $\chi_{\rm min}^2$ within the parameter space ($\varv_{\rm peak}, \Delta \varv$) and consider the error on the parameters of the fit such that $\Delta \chi^2 = \chi^2 - \chi_{\rm min}^2 = 1$, which corresponds to one standard deviation.
For each line we calculate the $\varv_{\rm peak}$ and $\Delta \varv$ of the best fit with relative errors, as a function of the disc inclination, for each of the different parameters listed in Table~\ref{tab:atomicparams}. We present our main results below. 
\begin{figure*}
	\centering
	\begin{subfigure}
        \centering
        \includegraphics[width=0.47\textwidth]{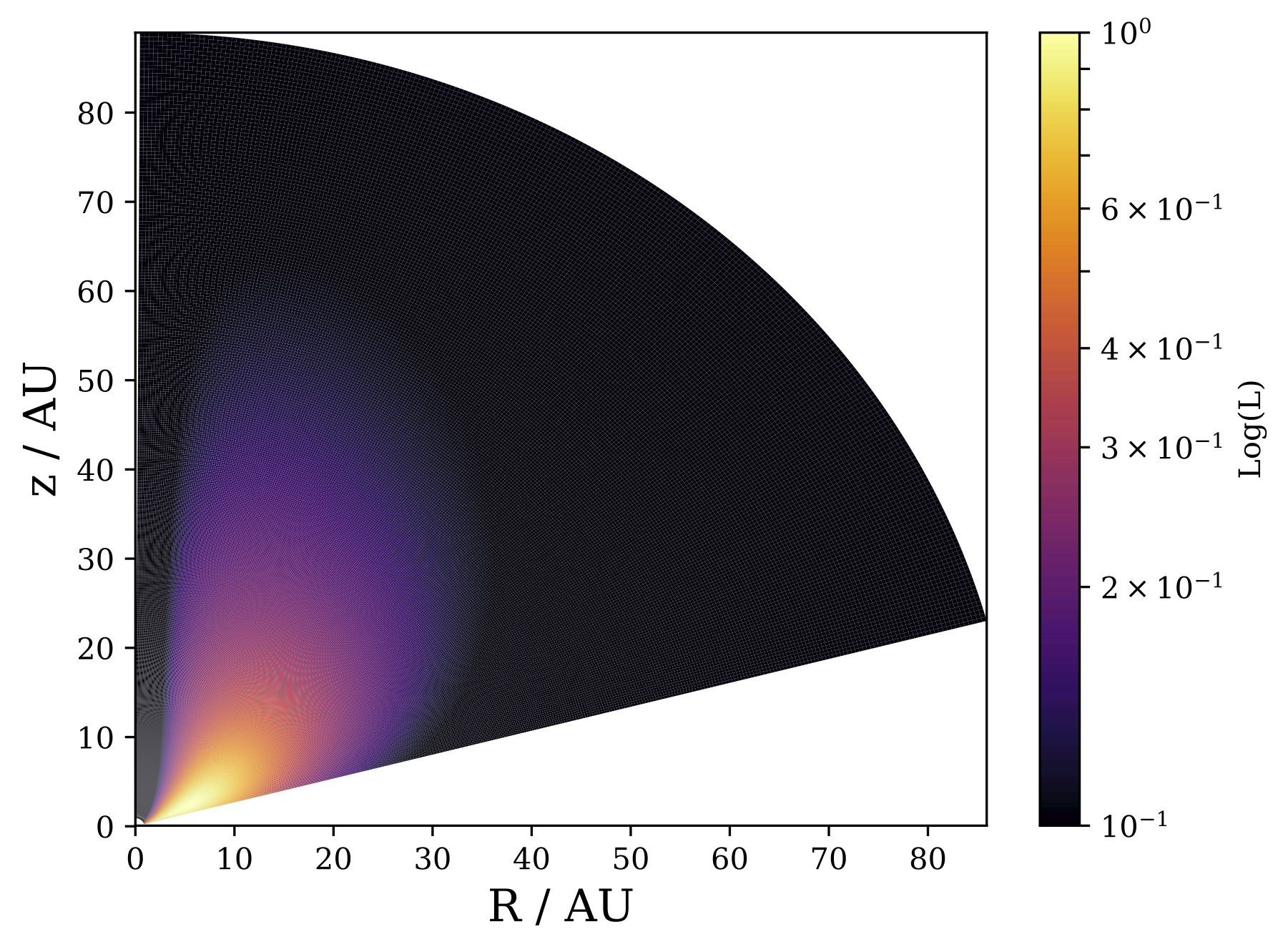}
        \hspace{0.5 cm}
        \includegraphics[width=0.47\textwidth]{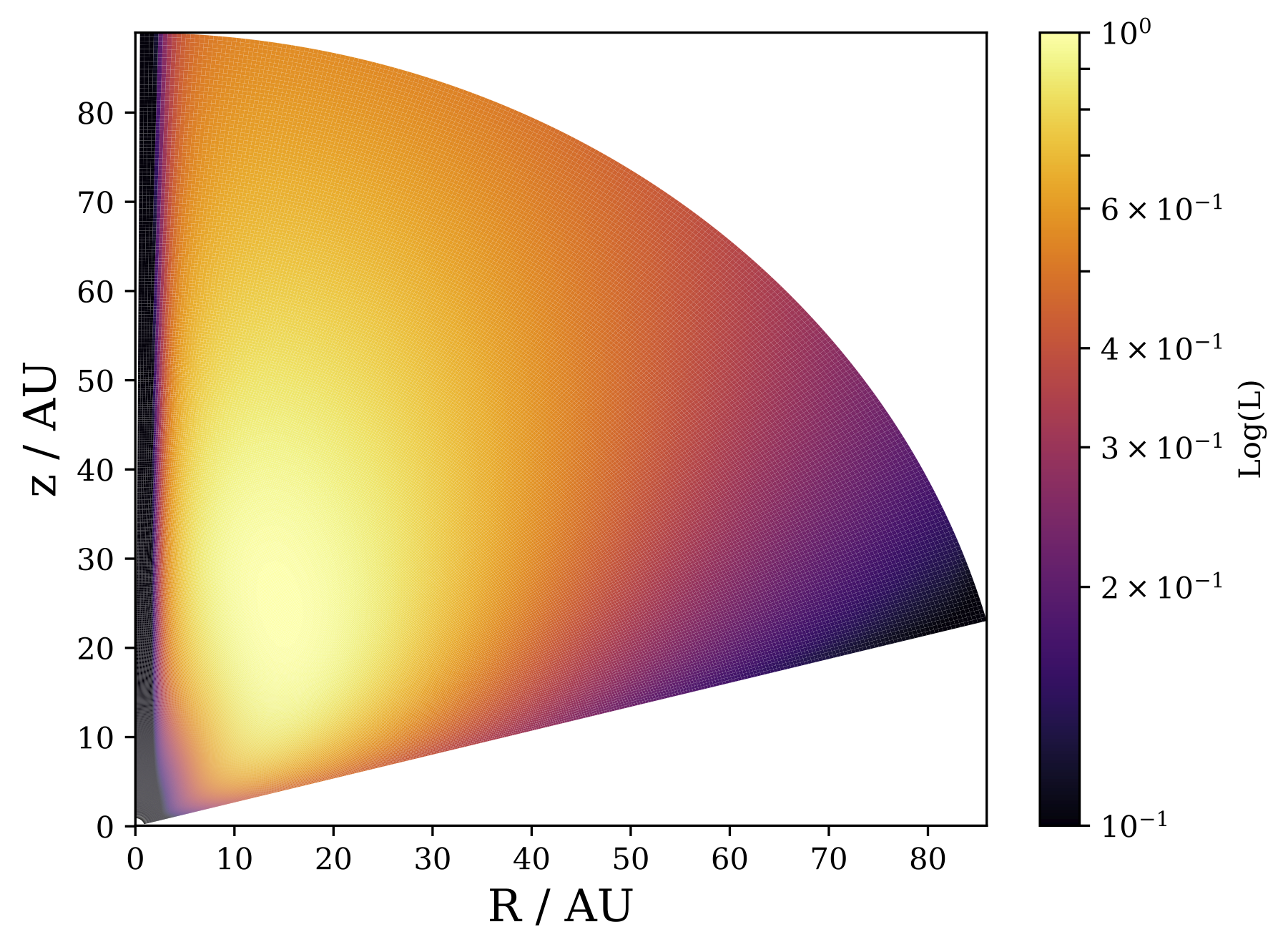}
        \caption{Volume emissivity in arbitrary units, for some of the lines considered in this work. Left panel: Emission map of the [Ne$\,${\sc{ii}}] line ($X_{\rm j}$=0.75). The [O$\,${\sc{i}}] line at 6300\AA~and [S$\,${\sc{ii}}] line at 4068\AA~present the same distribution. Tracers characterised by high critical densities mostly probe the inner region of the wind, closer to the disc mid-plane. Right panel: Emission map of the [S$\,${\sc{ii}}] line at 6716\AA\,($X_{\rm j}$=1). This sulphur line, characterised by a low critical density, traces mostly a region at large radii. Note that the spatial distribution of the emission shown here has been computed using data from hydrodynamic calculations (the same presented in Fig.~\ref{line_profile_comp} for the [Ne$\,${\sc{ii}}] line).}
        \label{flux_emission}
    \end{subfigure}
\end{figure*}

\subsection{What do the lines tell us?}
Blue-shifted emission lines are a clear sign of mass-loss from the disc, and blue-shifted [Ne$\,${\sc{ii}}] in particular is regarded as an unambiguous indicator of on-going photoevaporation \citep{2009ApJ...702..724P}. Here we consider the properties of the lines computed from our model in more detail. 

First, it is interesting to look at the spatial distribution of the line emission. Given density and velocity fields, we calculate the flux for each cell of the grid and plot the distribution of the volume emissivity \citep[following a similar approach to][]{2016MNRAS.460.3472E}. Fig.~\ref{flux_emission} shows two distinct spatial distribution of the emission between different tracers. It needs to be noted that the spatial distribution of the emission shown here has been computed using data from hydrodynamic calculations, for the purpose of the work. 
As we are considering forbidden lines, we see most of the emission only where the gas density drops below the critical density. The dependence of the emission on the density changes from squared to linear above the critical density, and as the density drops along the streamline and as a function of the disc radius, the emission is more often coming from near the critical density contour when you take the volume element into account.
As a consequence, lines with higher critical density mainly probe the densest regions, closer to the base of the flow. In the left panel of Fig.~\ref{flux_emission} we plot the [Ne$\,${\sc{ii}}] emission. For an isothermal wind the {\it relative} spatial distribution of the line flux is determined primarily by the critical density, so the [Ne$\,${\sc{ii}}] emission map is also representative of other high critical density lines such as [O$\,${\sc{i}}] 6300\AA~and [S$\,${\sc{ii}}] 4068\AA. The panel on the right-hand side of Fig.~\ref{flux_emission} shows the emission from the [S$\,${\sc{ii}}] line at 6716\AA, which is characteristic of low critical density tracers. Here the gas density is super-critical at the base of the flow, and drops below the critical density only at large distances along the streamlines. For low critical density lines we therefore see little or no emission from the inner region, and the line primarily traces the low-density outer regions of the wind.
\begin{figure}
	\centering
	\begin{subfigure}
        \centering
        \includegraphics[width=0.47\textwidth]{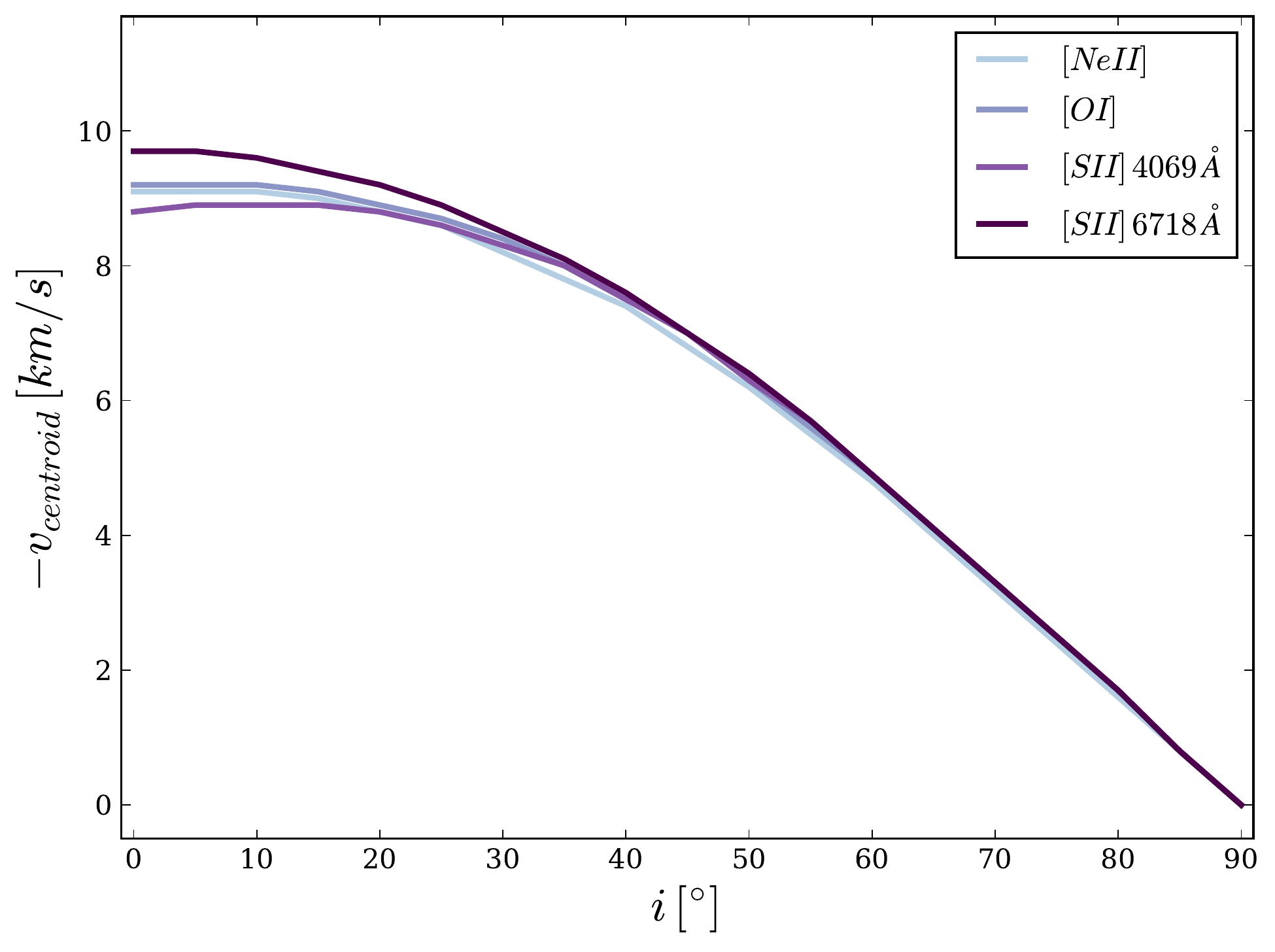}
    \end{subfigure}
    \begin{subfigure}
        \centering
        \includegraphics[width=0.47\textwidth]{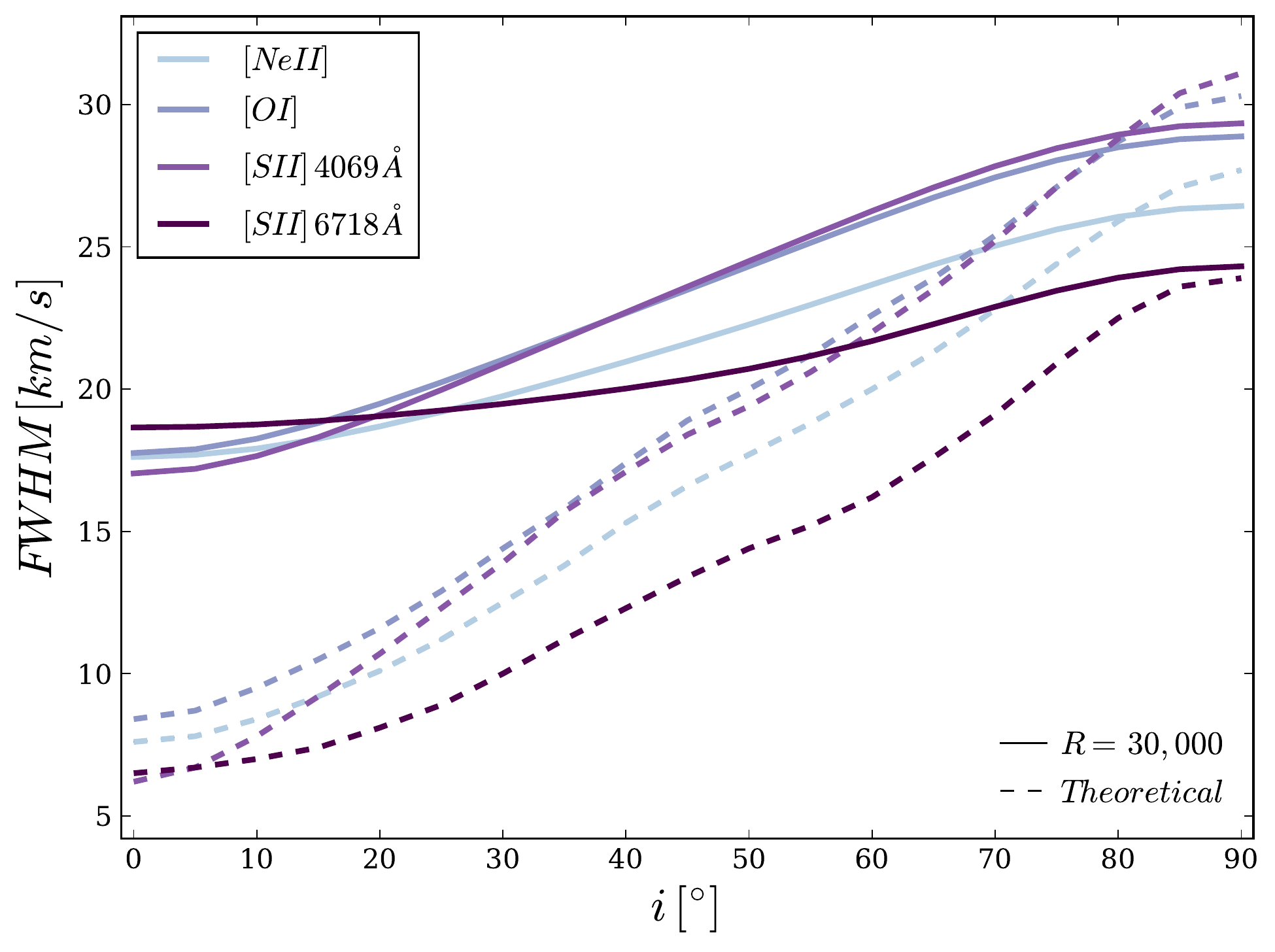}
    \end{subfigure}
    \caption{Blue-shifts (top) and line widths (bottom) as a function of the disc inclination, calculated for the tracers considered in this study (see Table~\ref{tab:atomicparams}). For these solutions, we consider a wind sound speed of $10$ km\,s$^{-1}$, with a density power law given by $b=1.50$ and normalised such as the mass loss rate within 25~au is $\dot{M} = 10^{-9} \rm{M}_{\odot}/\rm{yr}$. The solid lines identify lines degraded to a spectral resolution R=30$\,$000 and then fitted to a Gaussian profile, while the dashed lines (only in the bottom panel) are representative of the theoretical line profiles.}
    \label{fig:vpeak_fwhm_tracers}
\end{figure}

We study the line features, such as blue-shifts and width, calculating the velocity at the peak, the centroid velocity and the FWHM as functions of the disc inclination. We compute the emission lines for the species considered in this study and the results are presented in Fig.~\ref{fig:vpeak_fwhm_tracers}. The top and bottom panel show the centroid velocity and the FWHM, respectively. The lines are calculated from the theoretical solutions shown in Fig.~\ref{line_profile_comp} and normalised such as the mass loss rate within 25 au is $\dot{M} = 10^{-9} \rm{M}_{\odot}/\rm{yr}$. Furthermore, the lines are degraded to a spectral resolution R$=30\,000$ and fitted to a Gaussian function (solid lines). In the bottom panel, the width of both the theoretical (dashed lines) and the convolved (solid lines) line profiles is shown. 
Generally, our models are in good accordance with the values typically observed for these tracers, and all the lines show a similar trend. The blue-shift increases with decreasing disc inclination, reaching $\sim 9-10$ km\,s$^{-1}$ at low inclinations. Indeed, as shown in the top panel, there is very little variation of the centroid velocity between species. More interestingly, the line widths show quite significant variations among the tracers, suggesting possible hints on the diversity of the wind emitting region. As illustrated in the bottom panel of Fig.~\ref{fig:vpeak_fwhm_tracers}, tracers with a high critical density emit broader lines than tracers characterised by a low critical density. Indeed, transitions with a low critical density are probing regions at large radii and therefore we expect to see narrower lines. The discrepancy in the line widths is evident especially at high disc inclinations, as the emission at small radii and also with the highest contribution to the Keplerian velocity is more strongly suppressed (we refer the reader to Section~\ref{sec:low} for further discussion). It is also worth noting a systematic deviation in the line width at high disc inclinations of about $\sim 3$ km\,s$^{-1}$ between tracers with different critical densities.

Additionally, comparing the theoretical and the convolved widths, the discrepancy among the two distributions is remarkable. The convolution procedure largely affects the widths, broadening the lines, especially at lower inclinations. The range of values for the FWHM is also reduced significantly, by nearly half, resulting in shallow variations in the line widths. Moreover, fitting a single Gaussian profile to these complex, often asymmetric, lines results in loss of information especially for higher disc inclinations, for which we find that a Gaussian is not a particularly good fit. However, the main limitation comes from the spectral resolution, and a significant increase in the resolution will be required to distinguish between models clearly.
Nonetheless, we can divide our results into two categories: i) tracers with high critical densities (which include the neutral [O$\,${\sc{i}}] 6300\AA~line); and ii) tracers with low critical densities.

\subsubsection{High critical density}
\label{sec:high}
\begin{figure*}
    \centering
    \subfigure[]{\includegraphics[width=0.33\textwidth]{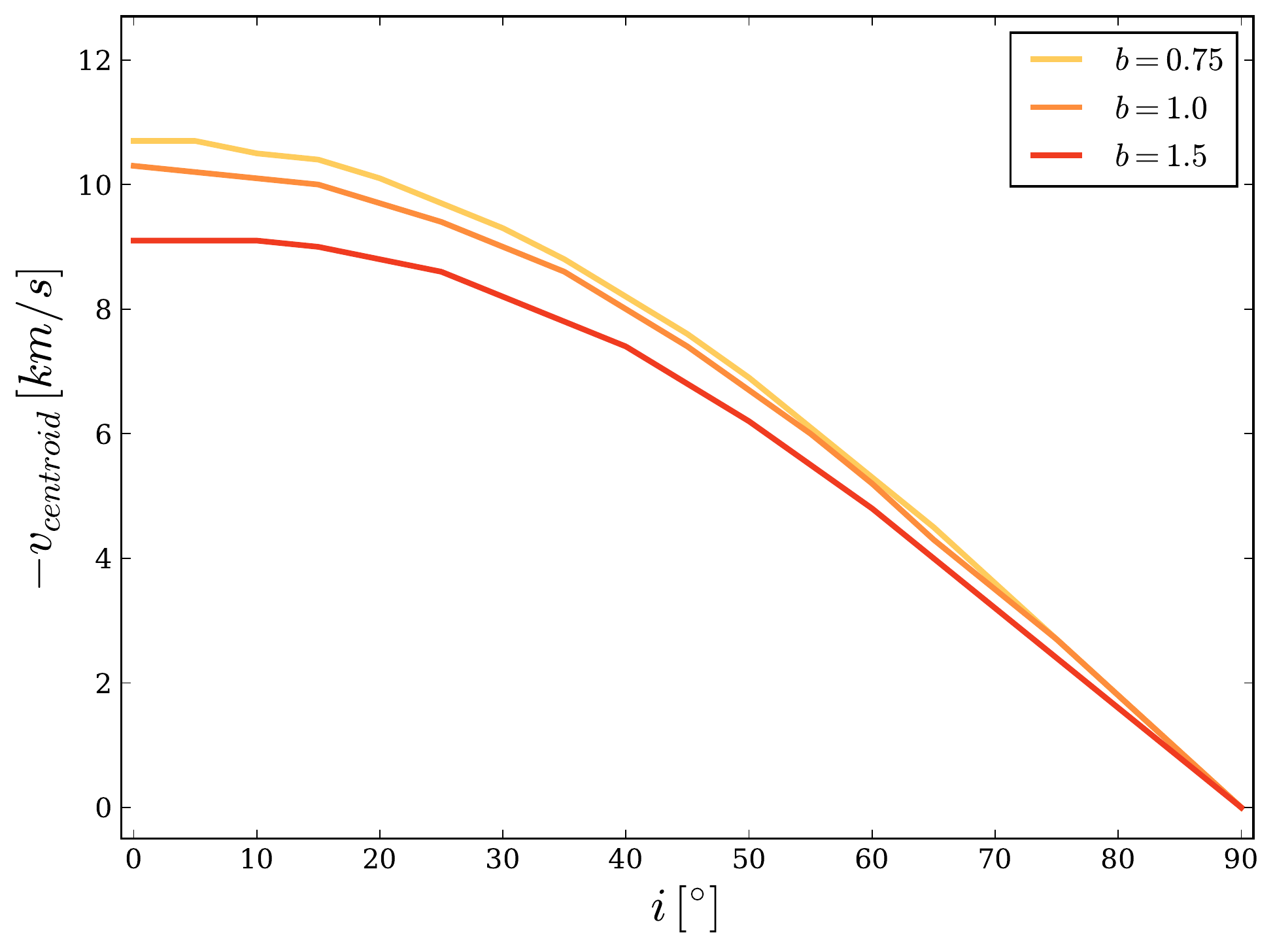}}
    \subfigure[]{\includegraphics[width=0.33\textwidth]{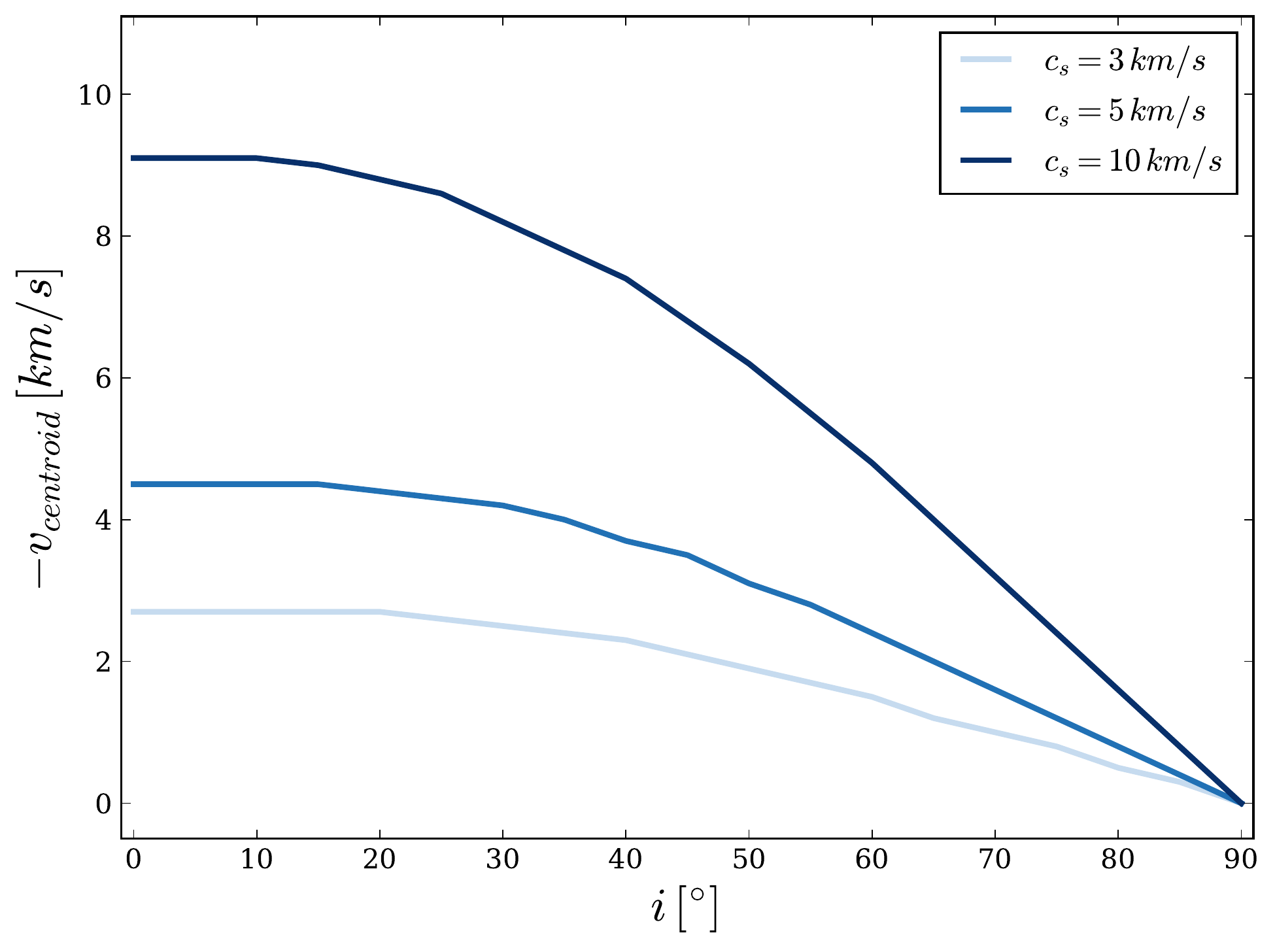}}
    \subfigure[]{\includegraphics[width=0.33\textwidth]{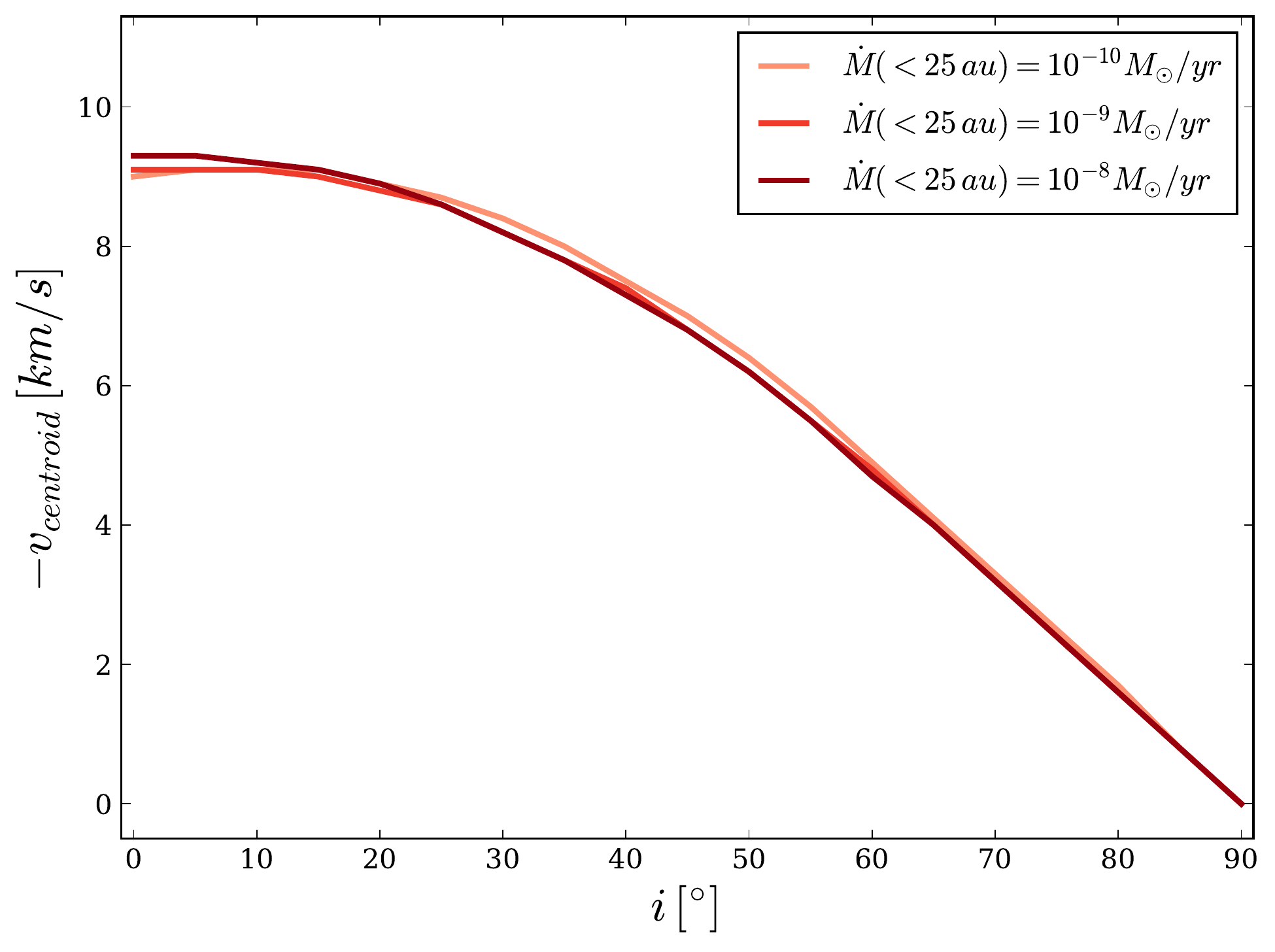}}
    \caption{Velocity of the centroid as a function of the disc inclination, calculated from the [Ne$\,${\sc{ii}}] line profiles shown in Fig.~\ref{line_profile_comp} at a spectral resolution R=30$\,$000. A Gaussian fit is also performed and then the blue-shifts are calculated for the parameters of the model listed in Table~\ref{tab:modelparams}. In the panels we present some of the results, where $b$, $c_{\mathrm s}$ and $\dot{M}$ are used as free parameters, respectively. Model with: (a) $c_{\mathrm s}=10$ km\,s$^{-1}$ and $\dot{M} = 10^{-9} \rm{M}_{\odot}/\rm{yr}$; (b) $b=1.5$ and $\dot{M} = 10^{-9} \rm{M}_{\odot}/\rm{yr}$; (c) $b=1.5$ and $c_{\mathrm s}=10$ km\,s$^{-1}$.}
    \label{fig:vcentr_incl}
\end{figure*}
\begin{figure*}
    \centering
    \subfigure[]{\includegraphics[width=0.33\textwidth]{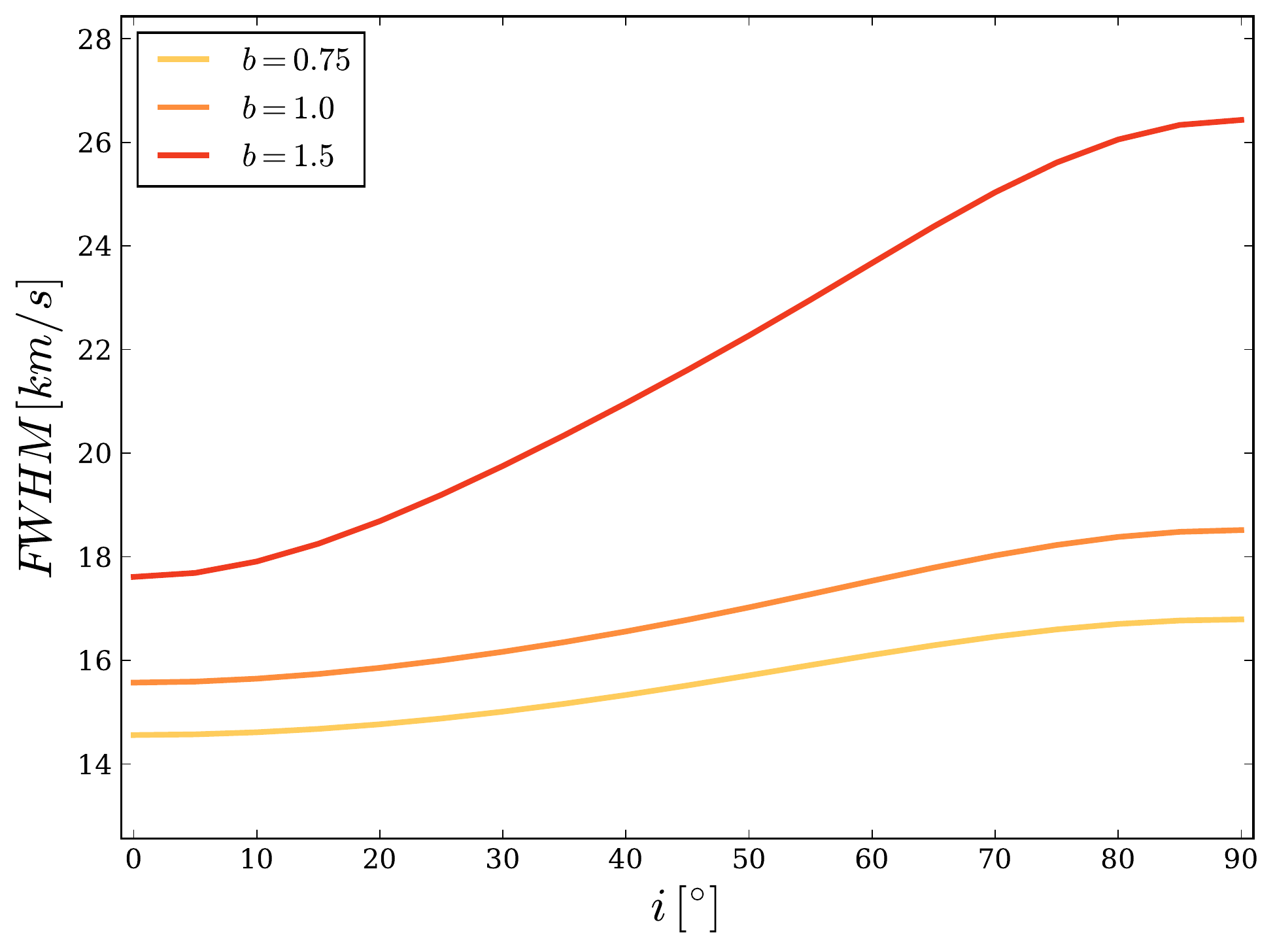}}
    \subfigure[]{\includegraphics[width=0.33\textwidth]{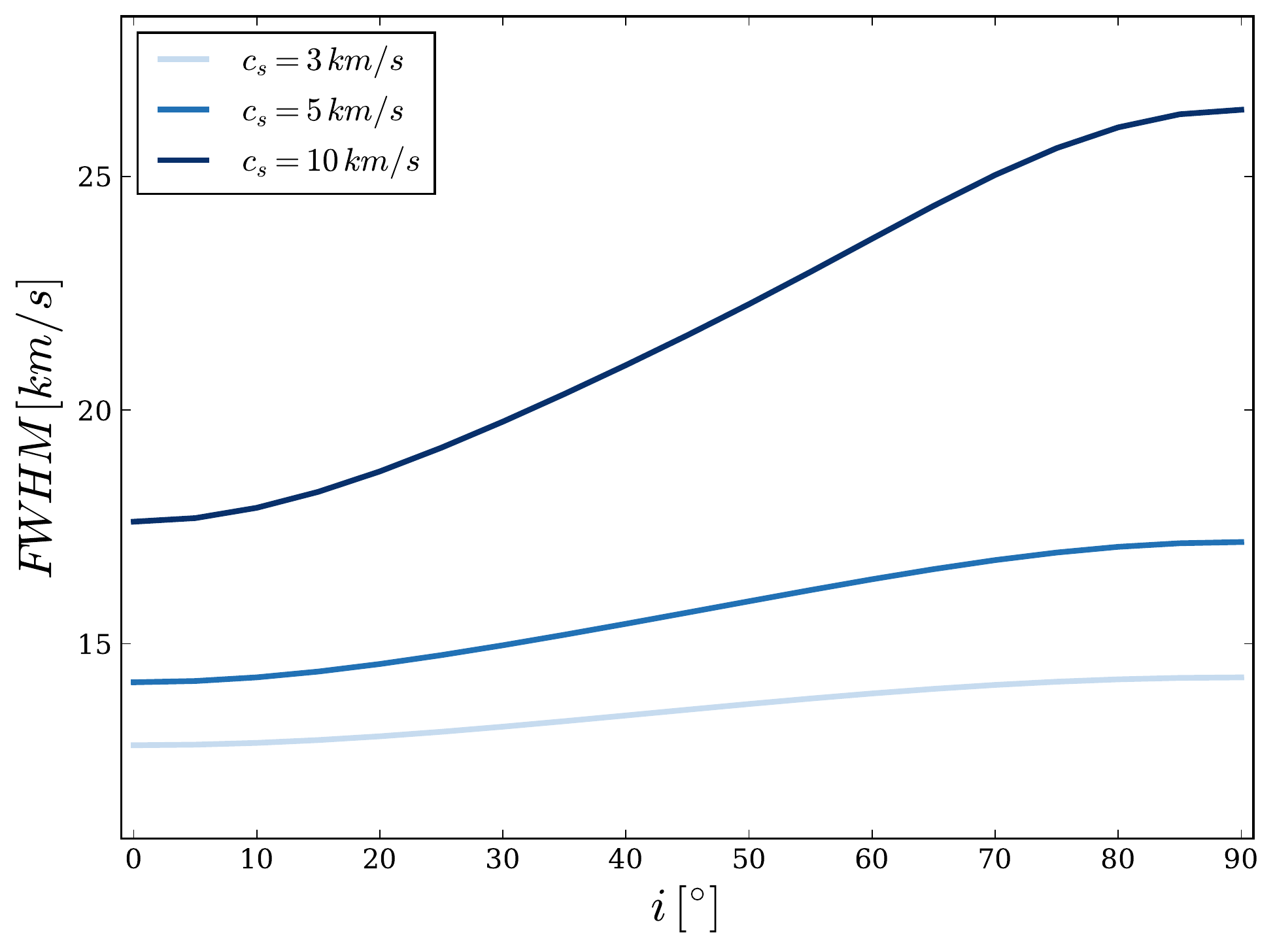}}
    \subfigure[]{\includegraphics[width=0.33\textwidth]{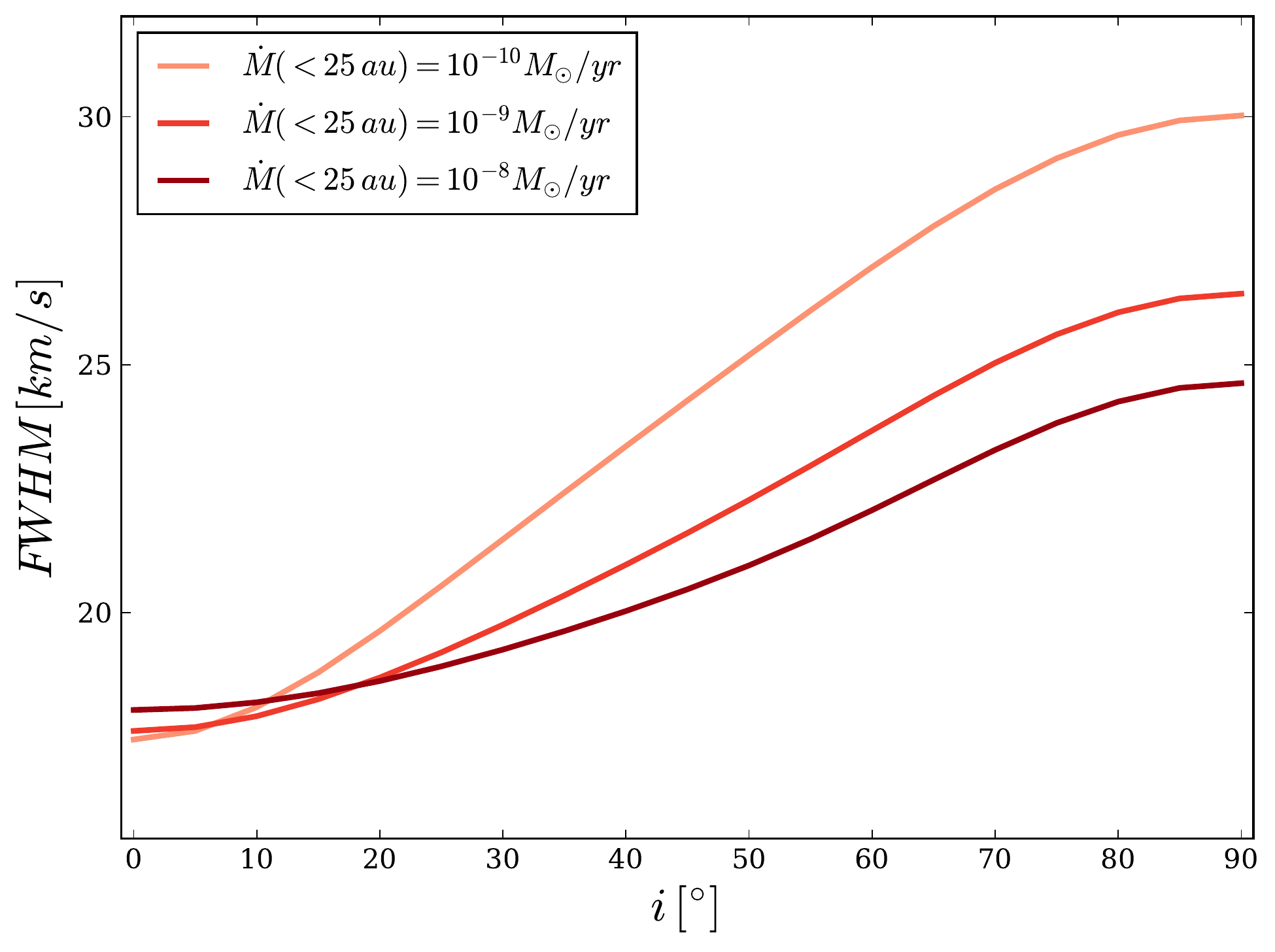}}
    \caption{FWHM as a function of the disc inclination, calculated from the [Ne$\,${\sc{ii}}] line profiles shown in Fig.~\ref{line_profile_comp} at a spectral resolution R=30$\,$000. A Gaussian fit is also performed and then the blue-shifts are calculated for the parameters of the model listed in Table~\ref{tab:modelparams}. In the panels we present some of the results, where $b$, $c_{\mathrm s}$ and $\dot{M}$ are used as free parameters, respectively. Model with: (a) $c_{\mathrm s}=10$ km\,s$^{-1}$ and $\dot{M} = 10^{-9} \rm{M}_{\odot}/\rm{yr}$; (b) $b=1.5$ and $\dot{M} = 10^{-9} \rm{M}_{\odot}/\rm{yr}$; (c) $b=1.5$ and $c_{\mathrm s}=10$ km\,s$^{-1}$.}
    \label{fig:fwhm_incl}
\end{figure*}
The [Ne$\,${\sc{ii}}] 12.81 $\mu$m line along with the [O$\,${\sc{i}}] 6300\AA~and [S$\,${\sc{ii}}] 4068/4076\AA~optical lines, are characterised by relatively high critical densities (see Table~\ref{tab:atomicparams}).
This is close to the typical gas density in the launching region of photoevaporative winds, and these lines are therefore excellent tracers of low-velocity thermal flows. They probe mostly the inner regions of the wind, as shown in the left panel of Fig.~\ref{flux_emission}, and are dominated by emission from near the base of the flow. Our results show that the normalized line profiles, particularly the line blue-shifts, for these different tracers are almost identical (see Fig.~\ref{fig:vpeak_fwhm_tracers}), and at $R \simeq 30\,000$ we do not expect to be able to detect small differences in the observables. Even in the case of the [O$\,${\sc{i}}] line for which the line flux is very sensitive to the ionization state, the normalized line profiles from our models are similar to the [Ne$\,${\sc{ii}}] lines.
For simplicity we therefore only present the general results for the [Ne$\,${\sc{ii}}] line. 

\paragraph{Line blue-shifts}
Fig.~\ref{fig:vcentr_incl} shows the centroid velocity as function of the disc inclination, for different values of the  parameters listed in Table~\ref{tab:modelparams}. The blue-shift increases with decreasing disc inclination, with differences in the values of the peak velocity more or less appreciable depending on the parameter considered. We generally find maximum observed blue-shifts of $\sim 9$--10\,km\,s$^{-1}$, at inclinations close to face-on. 

In panel~(a) of Fig.~\ref{fig:vcentr_incl} we show the centroid velocity as a function of the disc inclination, for different values of the parameter $b$. We fix the sound speed $c_{\mathrm s}=10$ km\,s$^{-1}$ and the mass loss rate $\dot{M} = 10^{-9} \rm{M}_{\odot}/\rm{yr}$. The blue-shifts are only marginally sensitive to the density profiles at the base of the wind. Nonetheless, our results show that a steeper density profile produces variations of a factor $\sim 1.25$ in the blue-shifts at low inclination angles. When the disc is highly inclined with respect to the line of sight (i.e. close to edge-on), there is no net blue-shift and the line profile is dominated by the Keplerian rotation of the wind base. 

We also study how the blue-shift changes with the gas sound speed, which can range from $c_{\mathrm s} \lesssim 3$ km\,s$^{-1}$ to $\gtrsim 10$ km\,s$^{-1}$ in X-ray and/or UV-driven photoevaporative winds. As we might expect, the observable blue-shifts are quite sensitive to the sound speed of the wind, as shown in panel~(b) of Fig.~\ref{fig:vcentr_incl}. Again the highest blue-shifts are seen close to face-on, and for high inclination angles the lines are dominated by the Keplerian rotation of the wind base. We plot the centroid velocity as a function of the disc inclination for different sound speeds and, as expected, we find that faster winds result in higher blue-shifts. However, the relationship between $c_{\mathrm s}$ and the peak blue-shift is not linear because of the additional line of sight components associated with the rotation of the wind base. 

We also compute the centroid velocity of the lines for different density normalization. The results are presented in panel~(c) of Fig.\ref{fig:vcentr_incl} and no appreciable differences in the blue-shifts are found. Increasing the base density pushes the critical density contour (above which emission is suppressed) slightly outwards, and the lack of any changes of the blue-shifts with $\dot{M}$ suggests that the centroid velocity is not sensitive to the critical density. 

\paragraph{Line widths}
More remarkable variations can be detected in the line broadening. In Fig.~\ref{fig:fwhm_incl}, we present the effect of the parameters considered in this study on the line widths. We observe a common trend, where the FWHM generally increases with increasing disc inclination. 
Panel~(a) shows the line widths as function of the disc inclination, again for different values of $b$. In contrast with the blue-shifts (see panel~(a) in Fig.~\ref{fig:vcentr_incl}), the FWHM is substantially affected by the density profile at the base of the wind. We can clearly appreciate the difference between each model, and shallower (lower $b$) density profiles results in narrower line-widths. This is somewhat expected, as steeper base density profiles (higher $b$) have higher densities at small radii, and therefore most of the emission comes from the innermost region (where the rotational velocity is larger), producing broader lines. On the contrary, less steep density power laws, which have still strong emission at larger radii, result in narrower lines. 

Like in the case of the blue-shifts, also the FWHM is more sensitive to the gas sound speed than any other parameter. As expected, faster winds result in broader line profiles, as shown in panel~(b) of Fig.~\ref{fig:fwhm_incl}. The observed widths span a range of values between $\sim 11$ and 22\,km\,s$^{-1}$. 

At last, it is also interesting to look at how the density normalization affects the FWHM as function of the disc inclination. Panel~(c) shows that greater $\dot{M}$ produce narrower lines, especially for higher disc inclinations. Increasing $\dot{M}$ moves the critical density contour to larger radii (where the rotational velocity is lower), and as a consequence the emission is more extended, generating narrower lines. 
\begin{figure*}
    \centering
    \includegraphics[width=\textwidth]{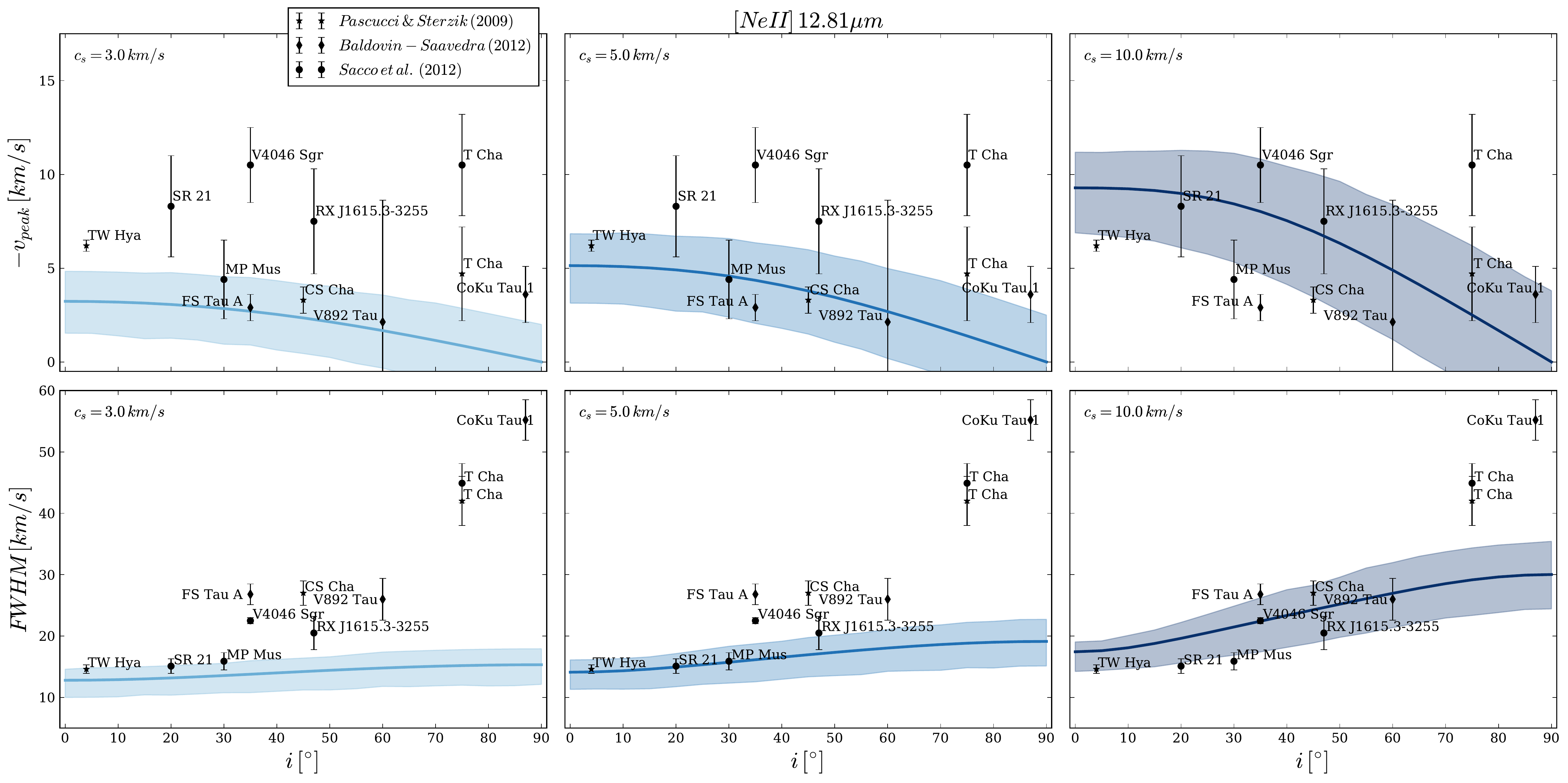}
    \caption{Peak velocity (top row) and FWHM (bottom row) as functions of the disc inclinations for the [Ne$\,${\sc{ii}}] line profile calculated at a spectral resolution R=30$\,$000. The plots in each column show the analytical solution at sound speeds 3, 5 and 10 \,km\,s$^{-1}$ respectively, and the colored bands represent the error on the parameters of the best fit, within one standard deviation. For comparison, we include in all panels resolved observations, taken from previous works (see legend). For this set of solutions, we fix $b=1.5$ and $\dot{M}=10^{-10} \rm{M}_{\odot}/\rm{yr}$.}
    \label{fig:soundspeed_data_NeII}
\end{figure*}

\paragraph{Comparison between the model and observations}
The [Ne$\,${\sc{ii}}] line at 12.81 micron is one of the best-studied of these lines observationally, and it is therefore a useful line for a comparison between our results and observations. In Fig.~\ref{fig:soundspeed_data_NeII} we plot the velocity at the peak and the width of the lines, in the top and bottom rows respectively, and we compare our models with the observed blue-shifts and FWHM for the [Ne$\,${\sc{ii}}] emission line for a sample of protoplanetary discs drawn from the literature. In each panel we compare the data with our analytical solutions at different sound speeds, 3, 5 and 10\,km\,s$^{-1}$. The shaded bands in the plots represent the errors on the parameters of the best fit, calculated from the $\chi^2$ test (see Section~\ref{sec:results}). The sample size is still modest, but, nonetheless, our model is generally consistent with the data within the error bars. In particular, the data primarily show blue-shifts in the range $\sim 5$--10\,km\,s$^{-1}$, and widths $\Delta \varv \sim 10-25$\,km\,s$^{-1}$, favouring models with higher sound speeds ($c_{\mathrm s} \gtrsim 10$ \,km\,s$^{-1}$); indeed it is difficult to reconcile most of the observed blue-shifts with a model that predicts $c_{\mathrm s}<5$\,km\,s$^{-1}$. It is worth noting that the line widths may be modestly under-estimated (by a couple of km/s) due to uncertainties associated with the volume wind modeled, as illustrated in Fig.~\ref{fig:line_convolfit} and Table~\ref{tab:valueslines}. Generally, the predicted line widths are thus somewhat narrower than the data in the sample, but a wind with sound speed $c_{\mathrm s} \sim 10$\,km\,s$^{-1}$ gives the best fit to the observed lines. This in turn points towards the higher wind temperatures expected for EUV-driven photoevaporation, rather than the somewhat cooler winds produced by X-ray or FUV irradiation. 
\begin{figure*}
    \centering
    \includegraphics[width=\textwidth]{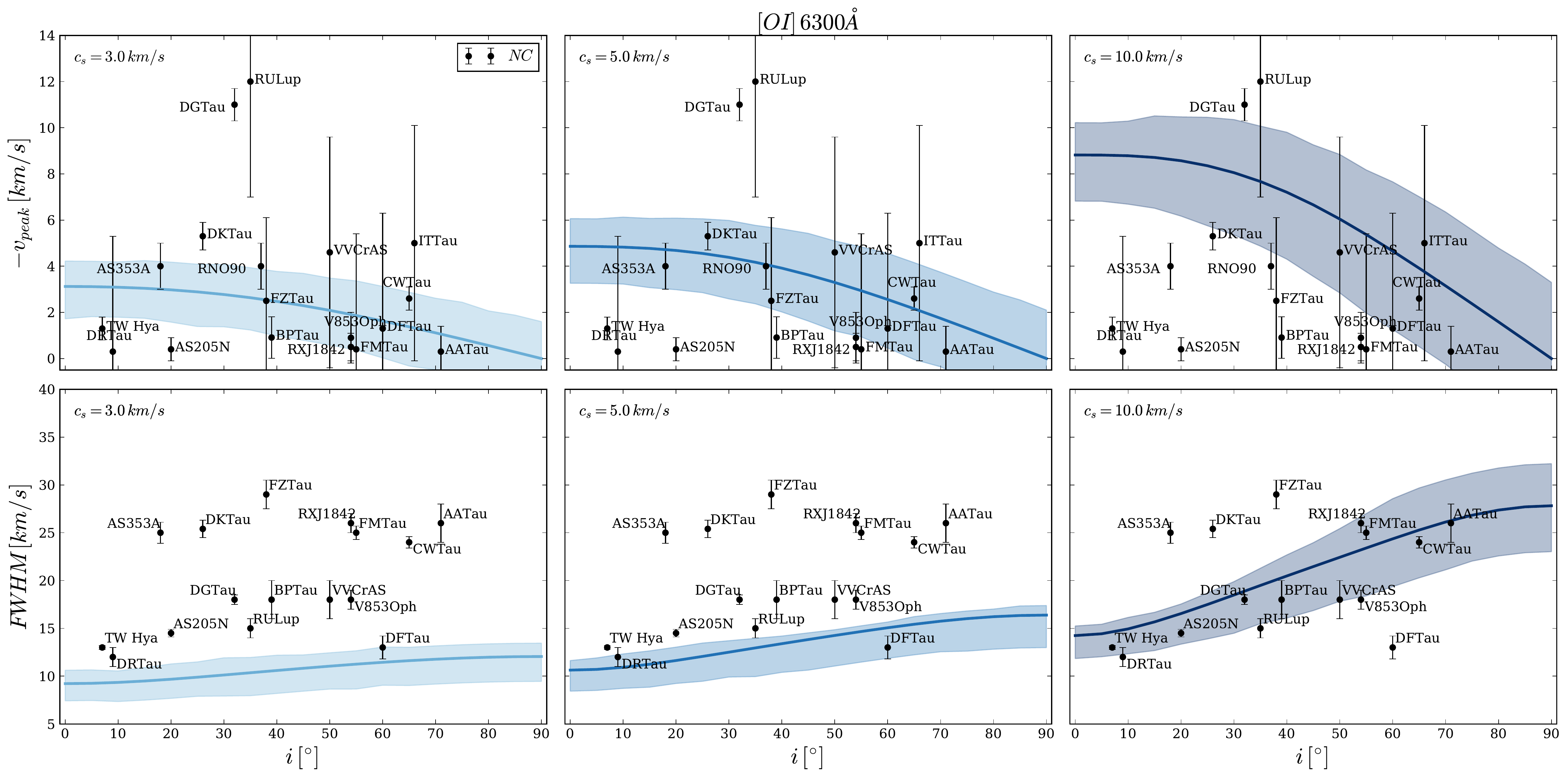}
    \caption{Peak velocity (top row) and FWHM (bottom row) of the [O$\,${\sc{i}}] line at 6300\AA. The line profiles are calculated at different disc inclinations and each column show the results at $c_{\mathrm s}=$3, 5 and 10\,km\,s$^{-1}$ respectively. We plot the results at a spectral resolution R=45\,000 (the resolution of Keck-HIRES and Magellan-MIKE observations of \citealt{2019ApJ...870...76B}). The black dots are the data taken from \citet{2019ApJ...870...76B} and only the narrow low velocity component is shown here. For these solutions, we fix $b=1.5$ and $\dot{M} = 10^{-9} \rm{M}_{\odot}/\rm{yr}$.}
    \label{fig:soundspeed_data_OI}
\end{figure*} 

Since the [O$\,${\sc{i}}] 6300\AA~is one of the strongest and most commonly observed lines, we also include it in our analysis and compare our theoretical model with the most recent high-resolution observations by \cite{2019ApJ...870...76B}. The [O$\,${\sc{i}}] line has a critical density comparable to the high-density ionized [Ne$\,${\sc{ii}}] line, but it is a neutral line.
The most interesting results here are the widths of the predicted [O$\,${\sc{i}}] lines, which are shown in the bottom row of Fig.~\ref{fig:soundspeed_data_OI}. As in the previous figure, each panel illustrates the results of the model derived for sound speeds of 3, 5 and 10\,km\,s$^{-1}$ respectively. For this set of solutions, the line profiles are convolved at a spectral resolution R=45\,000, the resolution of Keck-HIRES  and Magellan-MIKE observations of \citet{2019ApJ...870...76B}. The black dots are the narrow LVC data taken from the sample of \cite{2019ApJ...870...76B}. The line widths seem to favour higher sound speeds, and the model with $c_{\mathrm s}= 10$\,km\,s$^{-1}$ shows good broad agreement with the observations by \cite{2019ApJ...870...76B}. We caution, however, that these lines are not fully resolved (especially at lower inclinations and sound speeds), and this comparison is therefore quite sensitive to the assumed instrumental broadening.  We also note that our models fail to reproduce the small number of [O$\,${\sc{i}}] LVCs which show broad lines ($\sim 25-30$\,km\,s$^{-1}$) at small inclinations ($i < 45^{\circ}$). Such broad lines, especially at lower inclinations, presumably originate very close to the star (where the Keplerian velocity is larger), and appear to be inconsistent with a thermal wind origin. 
Along with the broadening, we also study the blue-shifts of the [O$\,${\sc{i}}] line, illustrated in the top row of Fig.~\ref{fig:soundspeed_data_OI}, where again we plot the narrow LVC data from \cite{2019ApJ...870...76B}. Again we see a strong dependence on the sound speed, but here the data have large error bars and do not allow us to discriminate strongly between models with different sound speeds. Lower sound speeds ($\sim 3-5$\,km\,s$^{-1}$) are weakly preferred, but it is also clear that a $10$\,km\,s$^{-1}$ wind is too hot to produce the observed [O$\,${\sc{i}}] line fluxes (due to thermal ionisation of the O$\,${\sc{i}}). We therefore conclude that a thermal wind model with sound speeds $\sim 3-5$\,km\,s$^{-1}$ fits the observed blue-shifts better than a higher velocity wind. 

Finally, we also consider the [S$\,${\sc{ii}}] line at 4068\AA,~comparing our models with the recent observations by \cite{2018ApJ...868...28F}. As in the case of the [O$\,${\sc{i}}], we only look at the narrow LVCs within their sample. The conclusion from our analysis is very similar to the results obtained for the [O$\,${\sc{i}}], in agreement with the findings of \cite{2018ApJ...868...28F}. The sample size is much smaller, but a $c_{\mathrm s} = 10$\,km\,s$^{-1}$ wind model provides the best fit to the observed [S$\,${\sc{ii}}] 4068\AA~lines. While the existence of neutral O$\,${\sc{i}} in such a wind is hard to explain (see the discussion in Section~\ref{sec:discussion}), the [S$\,${\sc{ii}}] emission is consistent with thermal ionization at $T\sim10^4$~K. However, the small sample size (only 7 objects) prevents us from drawing more detailed conclusions from the [S$\,${\sc{ii}}] data. 

\subsubsection{Low critical density}
\label{sec:low}
As an example of a low critical density tracer we consider the [S$\,${\sc{ii}}] line at 6716\AA. Here the gas density at the flow base exceeds the critical density by 2--4 orders of magnitude, so the line mostly probes the outer regions of the wind. This is shown in the right panel of Fig.~\ref{flux_emission}. For this reason, low critical density emission lines are not as useful tracers of thermal winds as the higher critical density lines considered above \citep{2004ApJ...607..890F}. In our calculations the [S$\,${\sc{ii}}] 6718\AA~line profile primarily probes the streamline topology at large radii ($R\gg R_{\mathrm g}$), and for such extended volume, differences between the self-similar and hydrodynamic streamline topologies are more apparent in the line profiles. 
The broken power-law profile of \citet{2004ApJ...607..890F} has $b=2.5$ in this region, so the density in our reference hydrodynamic model falls of much more steeply than in the self-similar solutions and we see substantial differences in the streamline topology.  Moreover, in the self-similar solutions the line flux diverges with increasing $R_{\mathrm {out}}$, and at fixed $\dot{M}$ the low critical density lines are therefore dominated by emission from the outer edge of the sampling volume. As a result we do not pursue the [S$\,${\sc{ii}}] 6718\AA~line further here. However, we also note that low critical density tracers are in general poor diagnostics of photoevaporative winds as, in addition to not probing the bulk of the flow, they are subject to significant ``contamination'' from emission in other parts of the star/disc/outflow system \citep[e.g.][]{1995ApJ...452..736H, 2004ApJ...607..890F, 2014A&A...569A...5N}.

\subsection{The line luminosity}
\begin{figure}
	\centering
	\includegraphics[width=0.47\textwidth,trim={0cm 0cm 0cm 0cm},clip]{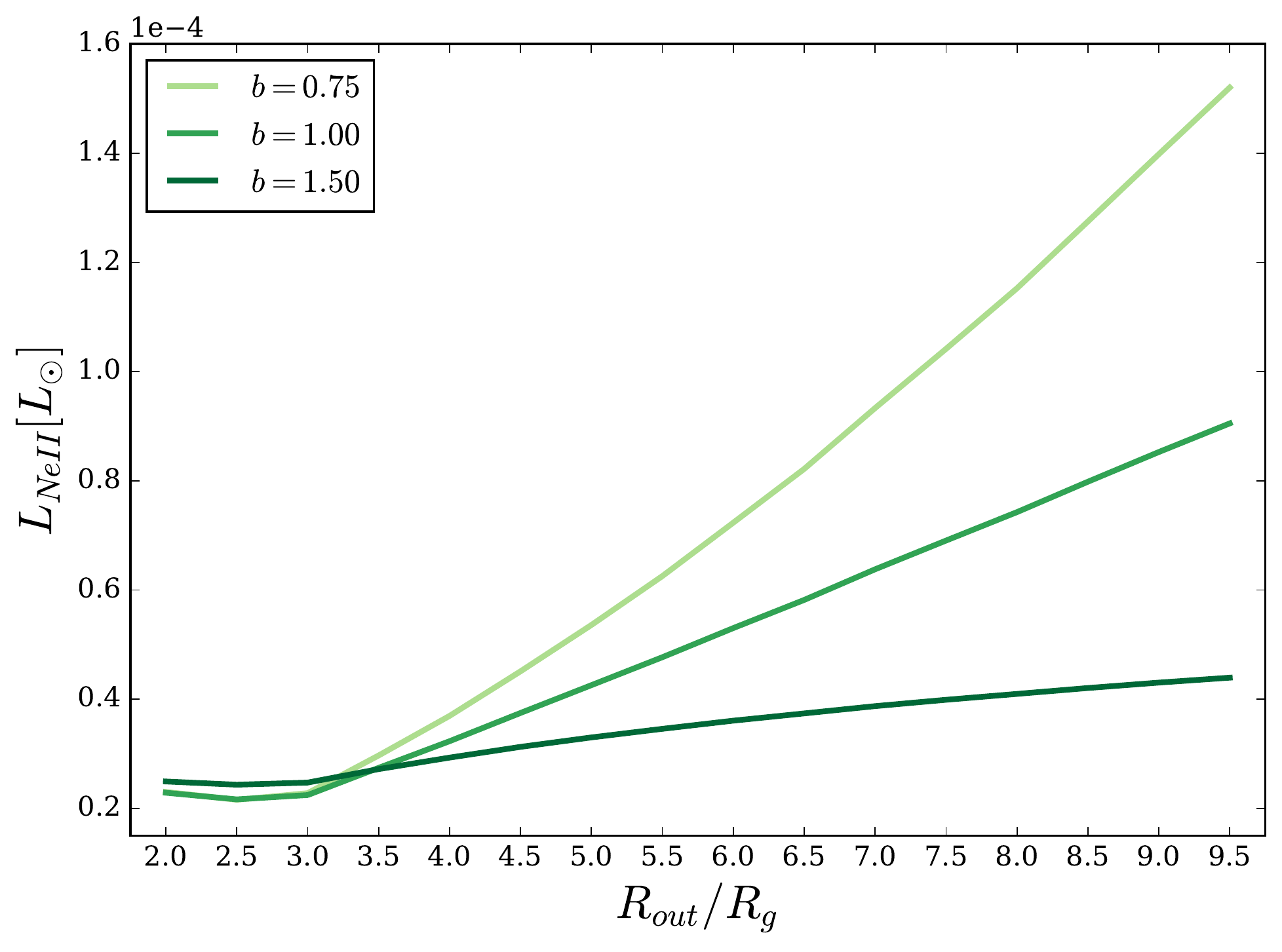}
	\caption{Total line luminosity as a function of the outer radius of the wind launch region, derived for the [Ne$\,${\sc{ii}}] line at $10$ km s$^{-1}$. The luminosity is calculated on a fix spherical grid and we sample the streamlines originating from a launch region of cylindrical extent $[R_{\rm in}, R_{\rm out}]$, where $R_{\rm in}=0.1\,{\rm R}_{\rm g}$ and $R_{\rm out}$ is equal to multiples of the gravitational radius. We fix the density normalisation as $\dot{M} = 10^{-9} \rm{M}_{\odot}/\rm{yr}$. We note that the line flux varies strongly with the density profile at the base of the wind.}
	\label{flux_radius}
\end{figure}

By assuming an isothermal wind with a constant ionization fraction, our model can also be used to calculate line luminosities. The absolute values of the line luminosities from our models are broadly consistent with the analytic calculations of \citet{2009ApJ...703.1203H}, and also previous radiative transfer calculations \citep{2010MNRAS.406.1553E, 2016MNRAS.460.3472E}. The integrated line luminosity can vary significantly even in cases where the (normalized) line profile is not strongly affected, and we find that the line luminosity depends primarily on the spatial extent of the emission (i.e., the outer radius of the wind launching region), and the base density profile. In order to show this, we keep the spherical grid volume fixed (as described in Section~\ref{sec:methods}) and considering different density profiles at the base of the flow, we sample the streamlines originating from a region of cylindrical extent $[R_{\rm in}, R_{\rm out}]$, where $R_{\rm in}=0.1\,{\rm R}_{\rm g}$ and $R_{\rm out}$ is equal to multiples of the gravitational radius. Fig.~\ref{flux_radius} shows the line luminosity as a function of $R_{\rm out}$ for different values of the power-law index $b$. For $n < n_{\mathrm {cr}}$ the integrated line luminosity scales as $L \propto n^2 R_{\mathrm {out}}^3$ (see Equation \ref{eq:lineprofile}), but we also note that at fixed $\dot{M}$ there is a degeneracy between the gas density and the radial extent of the wind. For fixed $\dot{M}$ we find that the line luminosity is an increasing function of $R_{\rm out}$, and that the gradient of this relation varies strongly with the density profile at the base of the wind, increasing much more rapidly for shallower density profiles (lower values of $b$). For very shallow base density profiles ($b\lesssim 1$) the line emission is extended across the disc, and the total luminosity diverges sharply to large radius. However, for steeper base density profiles ($b\gtrsim 1.5$), the bulk of the line flux originates at small radii, and the integrated flux is much less sensitive to the choice of $R_{\rm out}$.

We therefore conclude that the line luminosity primarily provides insight into the structure at the base of the flow (density normalization and profile), and the spatial extent of the flow, but when we consider line fluxes alone there is a degeneracy between these parameters. There are limited observational constraints which break this degeneracy, but we note that the spectro-astrometric observations of \citet{2011ApJ...736...13P} found that the [Ne\,{\sc ii}] emission from TW Hya originated at radii $\le 10$AU\,$=1.6\,{\rm R}_{\rm g}$. In the context of our model this points towards a relatively steep base density profile, and by comparison to Fig.\,\ref{flux_radius} we conclude that $b \gtrsim 1.5$ in this case.

\section{Discussion}
\label{sec:discussion}
Forbidden emission lines have long been used as a key diagnostic of protoplanetary disc winds, and several different tracers have been resolved in a large number of systems \citep[e.g.][]{1995ApJ...452..736H, 2007ApJ...663..383P, 2013ApJ...772...60R, 2014A&A...569A...5N}. Studying the velocity signatures of these lines has become the most common approach to deriving properties related to disc winds. However, observed spectra can be complex, and this makes it challenging to determine wind properties directly from the observations. Detailed models exist \citep[e.g.][]{2004ApJ...607..890F, 2010MNRAS.406.1553E, 2016MNRAS.460.3472E}, but - partly due to their complexity - there remains a disconnect between these models and observations, and many recent observational studies have preferred to take an empirical approach to understanding the line profiles. In this paper we have presented a new, analytical approach, using a simplified model to compute line profiles of forbidden lines, and then link observables to the physical properties of disc winds. Indeed, with our simpler approach we aim to gain new understanding of the physics of the problem, and from the results presented in Section~\ref{sec:results} we are able to infer some interesting new constraints.

\subsection{Observable properties of the wind}
In broad terms, we find that observable blue-shifts depend primarily on the disc inclination and wind sound speed, as shown in panel (b) of Fig.~\ref{fig:vcentr_incl}. By contrast, the blue-shift is relatively insensitive to the density structure in the wind ($b$ or $\dot{M}$) or the size of the emitting region, so if the inclination is known then the observed blue-shift can be used to estimate the wind sound speed (and therefore temperature).  
Further information about the sound speed of the wind can be derived from the widths, and they are consistent with what we find from the blue-shifts. The FWHM-inclination relation also varies strongly with the density profile, as we might expect. Both the base density profile and the density normalization sets the radius at which the density reaches its critical value: higher density values produce narrower lines, as the extent of super-critical material is wider. However, the observed line-widths are strongly affected by the spectral resolution, which broadens the lines. In particular, the effects of the finite resolution can be seen at the lowest inclinations, and in Figs.\,\ref{fig:soundspeed_data_NeII} \& \ref{fig:soundspeed_data_OI} the line widths at $i \lesssim 20^{\circ}$ are essentially just the instrumental resolution.

Our model allows us to calculate the line luminosities, albeit within the somewhat simplified context of an isothermal wind with a constant ionization fraction. We find a monotonic relationship between the integrated line luminosity and the outer radius of the wind launch region, which varies with the density normalisation (i.e. the parameter $b$). As a result, the line luminosity can provide further understanding of the structure at the base of the wind and the spatial extent of the flow. 
However, the obtained estimates are only absolute values of the line flux and they cannot break the degeneracy between the parameters. This is only a marginal caveat, as we are mostly interested in the line profiles. 

\subsection{Comparison with observations}
Our initial calculations show that the analytic model provides a good approximation to the results of more sophisticated calculations and, as shown in Section~\ref{sec:results}, our model reproduces the observed blue-shifts and widths of the forbidden emission lines well. The observed [Ne$\,${\sc{ii}}] 12.81$\mu$m line profiles in the literature typically have blue-shifts in the range $\simeq$5--10 km s$^{-1}$, and this generally points towards higher wind speeds. We require $c_s \gtrsim 10$\,km\,s$^{-1}$ for our models to reproduce the observed blue-shifts, and find that models with $c_s \lesssim 5$\,km\,s$^{-1}$ struggle to match the existing [Ne$\,${\sc{ii}}] observations.

The other forbidden line for which large numbers of high-quality line profiles exist is the neutral [O$\,${\sc{i}}] 6300\AA~line, and here the data lead us to a somewhat different conclusion. The [O$\,${\sc{i}}] line can have several different components, as discussed in Section~\ref{sec:intro}, and here we have focused on the narrow component of the LVC, which has previously been attributed to disc photoevaporation. Our models reproduce the observed line profiles relatively well, showing good agreement with the observed blue-shifts and widths. Our models are generally consistent with the [O$\,${\sc{i}}] line widths measured by \citet{2019ApJ...870...76B} (bottom panels of Fig.~\ref{fig:soundspeed_data_OI}), though the scatter in the observed widths does not allow to deduce any clear correlation between the FWHM and the disc inclination as predicted by our model (and the observed lines are only marginally resolved). 
We also note that the observed [O$\,${\sc{i}}] blue-shifts show significant scatter, especially at low inclination angles, but winds with $c_s \simeq$\,3--10\,km\,s$^{-1}$ are broadly consistent with the observed line profiles. While the observed blue-shifts are well described by models of cooler winds ($c_s \simeq$\,3--5\,km\,s$^{-1}$), the line widths are mainly in agreement with a 10\,km\,s$^{-1}$ disc wind. The contrast between these two results is not fully clear, but it could possibly hint at a wind origin different than thermal for these lines. We must also underline that the 6300\AA~line comes from neutral gas, and (as the ionization potential of O\,{\sc{i}} is 13.6eV) oxygen is thermally ionized at temperatures $\gtrsim$8000--10,000\,K. \citet{2004ApJ...607..890F} have shown that a fully ionized flow fails (by orders of magnitude) to reproduce the observed [O$\,${\sc{i}}] line luminosities, so we conclude that -- if they have a photoevaporative origin -- the [O$\,${\sc{i}}] 6300\AA~profiles are best-fit by sound speeds in the range $c_s \simeq$3--5\,km\,s$^{-1}$. 
We also distinguish a small population of low-inclination discs with large [O$\,${\sc{i}}] widths, $\Delta \sim 25-30$\,km\,s$^{-1}$ (Fig.~\ref{fig:soundspeed_data_OI}). This appears inconsistent with a thermal wind origin, and suggests that some other mechanism is responsible for the observed emission in these cases.
We also note the few sources which show narrow [O$\,${\sc{i}}] lines which have no measurable blue-shift. This is suggestive not only of a non-thermal origin, but that in some sources this line does not trace a wind at all. 

Finally, is it worth bearing in mind that the line profiles from our models are usually asymmetric and/or double-peaked, and are significantly more complex than the Gaussian profiles used in the empirical fitting procedure of \cite{2019ApJ...870...76B}. Indeed, when we look at the composite line profiles (both broad and narrow components) in Fig.~15 of \cite{2019ApJ...870...76B}, there is a striking similarity between some of the observed profiles and our ``raw'' model lines (shown in Fig.\,\ref{line_profile_comp}), especially for discs close to a face-on configuration (such as DR Tau or AS 205N). This suggests that a more sophisticated analysis of the observed line profiles should be carried out, and that fitting physically-motivated models (such as those presetented here) to these data may yield greater insight than a Gaussian decomposition alone.

Thus far we have considered our results in the abstract, but it is instructive at this point to note again that the wind sound speed is a key distinction between different ``flavours'' of disc photoevaporation. EUV-driven winds are fully ionized and isothermal, with $c_s \simeq$10\,km\,s$^{-1}$ \citep{1994ApJ...428..654H, 2004ApJ...607..890F}, while X-ray heating typically results in lower temperatures (3000--5000 K) and some departure from isothermality \citep[e.g.,][though we note that most recent models by \citealp{2019MNRAS.487..691P} predict near-isothermal winds in the flow region]{2010MNRAS.401.1415O}\footnote{FUV-heating results in lower temperatures still (typically 500--1000K, \citealt{2009ApJ...690.1539G}), and while FUV winds can drive significant mass-loss at larger radii they cannot explain the flow velocities or ionization states of the forbidden lines considered here.}. Previous studies have come to different conclusions about the origin of these line profiles \citep{2010MNRAS.406.1553E, 2013ApJ...772...60R, 2016MNRAS.460.3472E, 2018ApJ...868...28F, 2019ApJ...870...76B}. We find that it is difficult to reproduce both the [Ne$\,${\sc{ii}}] and [O$\,${\sc{i}}] lines with a single, isothermal flow, and that the [Ne$\,${\sc{ii}}] emission appears to come from hotter, higher-velocity gas than the [O$\,${\sc{i}}]. It may be that these lines probe different regions of a (strongly) non-isothermal photoevaporative wind, or that the flow has a non-thermal origin (e.g. a magnetically-launched wind), but more sophisticated models -- and further observations -- are needed to resolve this question.

We also stress that observations of discs at low inclination angles remain the best way to study low-velocity winds empirically.  At high inclinations ($i\gtrsim60^{\circ}$) the line profiles are dominated by rotation, making it difficult to separate any contribution from bound disc material from the emission arising in the wind. By contrast, blue-shifted emission at low inclination angles ($i \lesssim 45^{\circ}$) is an unambiguous wind signature, and our results show that these line profiles can provide a direct probe of the underlying wind structure.

Finally, we note that the typical spectral resolution of these observations -- $R \sim $30,000--40,000 -- means that many of these lines are only marginally resolved. Determining line parameters by fitting Gaussian line profiles also introduces further uncertainties, as our models (and others) have shown that winds are expected to result in line profiles that are not well fit by a Gaussian profile. These factors limit our ability to discriminate between models using current data, and our results show that significantly higher spectral resolution ($R \gtrsim 100,000$) is required if we are to use these line profiles to measure the properties of disc winds in greater detail.

\section{Conclusions}
\label{sec:concs}
We have presented a new approach to calculate observable diagnostics of photoevaporative disc winds. We use an analytic model of an isothermal disc wind, coupled to a simplified emissivity, to create a straight-forward and fast method to derive forbidden emission line profiles and related observables. The simplicity of this approach allows us to explore a wide range of parameters, and to investigate in detail how the various physical parameters affect the emission lines. We compute emission lines for typical tracers of ionized thermal winds, focusing on the [Ne$\,${\sc{ii}}] 12.81 $\mu$m, [S$\,${\sc{ii}}] 4068/4076\AA~and [O$\,${\sc{i}}] 6300\AA~lines. Our model successfully reproduces previous results, and we show that these lines have blue-shifts of $\lesssim 10$\,km\,s$^{-1}$ which decrease with increasing disc inclination. The line widths are typically $\sim 15-30$\,km\,s$^{-1}$, and increase with the disc inclination. 
At high inclination the line profiles are dominated by rotation, but observations at low inclination angles can probe the wind structure directly.
We find that the observed line blue-shift primarily traces the sound speed of the underlying flow, but it is difficult to reproduce the observations of both ionized and neutral lines with a single model. We find that the observed [Ne$\,${\sc{ii}}] 12.81 $\mu$m line profiles favour higher sound speeds ($c_s \gtrsim 10$\,km\,s$^{-1}$), as expected for an EUV-driven photoevaporative wind, while the [O$\,${\sc{i}}] 6300\AA~lines traces cooler, lower-velocity gas ($c_s \simeq 3$--5\,km\,s$^{-1}$), as expected for either X-ray-driven photoevaporation or for a magnetically-launched wind. It is possible that multiple launching processes may be at work simultaneously, but it seems more likely that these lines trace different components of a multi-phase wind (such as that proposed by \citealp{2016ApJ...818..152B}). Finally, we note that observations of these low-velocity winds remain limited by instrumental resolution, and that measuring their properties in greater detail is likely to require new observations at significantly higher spectral resolution ($\lambda/\Delta \lambda \gtrsim 100\, 000$). 

\section*{Acknowledgments}
We are grateful to an anonymous referee for constructive suggestions and a detailed report that improved our paper.
We thank Antonella Natta, Ilaria Pascucci and Andrea Banzatti for fruitful discussions. We also thank our colleagues Rebecca Nealon and Enrico Ragusa for useful comments on a preliminary draft. G.B. acknowledges support from the University of Leicester through a College of Science and Engineering PhD studentship. 
This project has received funding from the European Research Council (ERC) under the European Union's Horizon 2020 research and innovation program (grant agreement No 681601). This project has been carried out as part of the European Union's Horizon 2020 research and innovation programme under the Marie Sk\l odowska-Curie grant agreement No 823823 (DUSTBUSTERS).
This work was supported by the STFC Consolidated Grant ST/S000623/1.
This research used the ALICE High Performance Computing Facility at the University of Leicester and the DiRAC \textit{Complexity} system, operated by the University of Leicester IT Services, which forms part of the STFC DiRAC HPC Facility.


\section*{Data Availability}
The data underlying this article will be shared on reasonable request to the corresponding author.



\bibliographystyle{mnras}
\bibliography{mybibliography.bib}

\begin{thebibliography}{}
\makeatletter
\relax
\def\mn@urlcharsother{\let\do\@makeother \do\$\do\&\do\#\do\^\do\_\do\%\do\~}
\def\mn@doi{\begingroup\mn@urlcharsother \@ifnextchar [ {\mn@doi@}
  {\mn@doi@[]}}
\def\mn@doi@[#1]#2{\def\@tempa{#1}\ifx\@tempa\@empty \href
  {http://dx.doi.org/#2} {doi:#2}\else \href {http://dx.doi.org/#2} {#1}\fi
  \endgroup}
\def\mn@eprint#1#2{\mn@eprint@#1:#2::\@nil}
\def\mn@eprint@arXiv#1{\href {http://arxiv.org/abs/#1} {{\tt arXiv:#1}}}
\def\mn@eprint@dblp#1{\href {http://dblp.uni-trier.de/rec/bibtex/#1.xml}
  {dblp:#1}}
\def\mn@eprint@#1:#2:#3:#4\@nil{\def\@tempa {#1}\def\@tempb {#2}\def\@tempc
  {#3}\ifx \@tempc \@empty \let \@tempc \@tempb \let \@tempb \@tempa \fi \ifx
  \@tempb \@empty \def\@tempb {arXiv}\fi \@ifundefined
  {mn@eprint@\@tempb}{\@tempb:\@tempc}{\expandafter \expandafter \csname
  mn@eprint@\@tempb\endcsname \expandafter{\@tempc}}}

\bibitem[\protect\citeauthoryear{{Adams}, {Hollenbach}, {Laughlin}  \&
  {Gorti}}{{Adams} et~al.}{2004}]{2004ApJ...611..360A}
{Adams} F.~C.,  {Hollenbach} D.,  {Laughlin} G.,   {Gorti} U.,  2004, \mn@doi
  [\apj] {10.1086/421989}, \href
  {https://ui.adsabs.harvard.edu/abs/2004ApJ...611..360A} {611, 360}

\bibitem[\protect\citeauthoryear{{Alexander}}{{Alexander}}{2008}]{2008MNRAS.391L..64A}
{Alexander} R.~D.,  2008, \mn@doi [\mnras] {10.1111/j.1745-3933.2008.00556.x},
  \href {https://ui.adsabs.harvard.edu/abs/2008MNRAS.391L..64A} {391, L64}

\bibitem[\protect\citeauthoryear{{Alexander}, {Clarke}  \&
  {Pringle}}{{Alexander} et~al.}{2006a}]{2006MNRAS.369..216A}
{Alexander} R.~D.,  {Clarke} C.~J.,   {Pringle} J.~E.,  2006a, \mn@doi [\mnras]
  {10.1111/j.1365-2966.2006.10293.x}, \href
  {https://ui.adsabs.harvard.edu/abs/2006MNRAS.369..216A} {369, 216}

\bibitem[\protect\citeauthoryear{{Alexander}, {Clarke}  \&
  {Pringle}}{{Alexander} et~al.}{2006b}]{2006MNRAS.369..229A}
{Alexander} R.~D.,  {Clarke} C.~J.,   {Pringle} J.~E.,  2006b, \mn@doi [\mnras]
  {10.1111/j.1365-2966.2006.10294.x}, \href
  {https://ui.adsabs.harvard.edu/abs/2006MNRAS.369..229A} {369, 229}

\bibitem[\protect\citeauthoryear{{Alexander}, {Pascucci}, {Andrews}, {Armitage}
   \& {Cieza}}{{Alexander} et~al.}{2014}]{2014prpl.conf..475A}
{Alexander} R.,  {Pascucci} I.,  {Andrews} S.,  {Armitage} P.,   {Cieza} L.,
  2014, Protostars and Planets VI, \href
  {https://doi.org/10.2458/azu_uapress_9780816531240-ch021} {pp 475--496}

\bibitem[\protect\citeauthoryear{{Andrews} \& {Williams}}{{Andrews} \&
  {Williams}}{2005}]{2005ApJ...631.1134A}
{Andrews} S.~M.,  {Williams} J.~P.,  2005, \mn@doi [\apj] {10.1086/432712},
  \href {https://ui.adsabs.harvard.edu/abs/2005ApJ...631.1134A} {631, 1134}

\bibitem[\protect\citeauthoryear{{Armitage}}{{Armitage}}{2011}]{2011ARA&A..49..195A}
{Armitage} P.~J.,  2011, \mn@doi [\araa] {10.1146/annurev-astro-081710-102521},
  \href {https://ui.adsabs.harvard.edu/abs/2011ARA%26A..49..195A} {49, 195}

\bibitem[\protect\citeauthoryear{{Asplund}, {Grevesse}, {Sauval}  \&
  {Scott}}{{Asplund} et~al.}{2009}]{2009ARA&A..47..481A}
{Asplund} M.,  {Grevesse} N.,  {Sauval} A.~J.,   {Scott} P.,  2009, \mn@doi
  [\araa] {10.1146/annurev.astro.46.060407.145222}, \href
  {http://adsabs.harvard.edu/abs/2009ARA%26A..47..481A} {47, 481}

\bibitem[\protect\citeauthoryear{{Bai}, {Ye}, {Goodman}  \& {Yuan}}{{Bai}
  et~al.}{2016}]{2016ApJ...818..152B}
{Bai} X.-N.,  {Ye} J.,  {Goodman} J.,   {Yuan} F.,  2016, \mn@doi [\apj]
  {10.3847/0004-637X/818/2/152}, \href
  {https://ui.adsabs.harvard.edu/abs/2016ApJ...818..152B} {818, 152}

\bibitem[\protect\citeauthoryear{{Banzatti}, {Pascucci}, {Edwards}, {Fang},
  {Gorti}  \& {Flock}}{{Banzatti} et~al.}{2019}]{2019ApJ...870...76B}
{Banzatti} A.,  {Pascucci} I.,  {Edwards} S.,  {Fang} M.,  {Gorti} U.,
  {Flock} M.,  2019, \mn@doi [\apj] {10.3847/1538-4357/aaf1aa}, \href
  {https://ui.adsabs.harvard.edu/abs/2019ApJ...870...76B} {870, 76}

\bibitem[\protect\citeauthoryear{{Begelman}, {McKee}  \& {Shields}}{{Begelman}
  et~al.}{1983}]{1983ApJ...271...70B}
{Begelman} M.~C.,  {McKee} C.~F.,   {Shields} G.~A.,  1983, \mn@doi [\apj]
  {10.1086/161178}, \href
  {https://ui.adsabs.harvard.edu/abs/1983ApJ...271...70B} {271, 70}

\bibitem[\protect\citeauthoryear{{Clarke} \& {Alexander}}{{Clarke} \&
  {Alexander}}{2016}]{2016MNRAS.460.3044C}
{Clarke} C.~J.,  {Alexander} R.~D.,  2016, \mn@doi [\mnras]
  {10.1093/mnras/stw1178}, \href
  {https://ui.adsabs.harvard.edu/abs/2016MNRAS.460.3044C} {460, 3044}

\bibitem[\protect\citeauthoryear{{Clarke}, {Gendrin}  \& {Sotomayor}}{{Clarke}
  et~al.}{2001}]{2001MNRAS.328..485C}
{Clarke} C.~J.,  {Gendrin} A.,   {Sotomayor} M.,  2001, \mn@doi [\mnras]
  {10.1046/j.1365-8711.2001.04891.x}, \href
  {https://ui.adsabs.harvard.edu/abs/2001MNRAS.328..485C} {328, 485}

\bibitem[\protect\citeauthoryear{{Draine}}{{Draine}}{2011}]{2011piim.book.....D}
{Draine} B.~T.,  2011, {Physics of the Interstellar and Intergalactic Medium}.
Princeton University Press

\bibitem[\protect\citeauthoryear{{Dullemond}, {Hollenbach}, {Kamp}  \&
  {D'Alessio}}{{Dullemond} et~al.}{2007}]{2007prpl.conf..555D}
{Dullemond} C.~P.,  {Hollenbach} D.,  {Kamp} I.,   {D'Alessio} P.,  2007,
  Protostars and Planets V, \href
  {https://ui.adsabs.harvard.edu/abs/2007prpl.conf..555D} {p.~555}

\bibitem[\protect\citeauthoryear{{Ercolano} \& {Owen}}{{Ercolano} \&
  {Owen}}{2010}]{2010MNRAS.406.1553E}
{Ercolano} B.,  {Owen} J.~E.,  2010, \mn@doi [\mnras]
  {10.1111/j.1365-2966.2010.16798.x}, \href
  {https://ui.adsabs.harvard.edu/abs/2010MNRAS.406.1553E} {406, 1553}

\bibitem[\protect\citeauthoryear{{Ercolano} \& {Owen}}{{Ercolano} \&
  {Owen}}{2016}]{2016MNRAS.460.3472E}
{Ercolano} B.,  {Owen} J.~E.,  2016, \mn@doi [\mnras] {10.1093/mnras/stw1179},
  \href {https://ui.adsabs.harvard.edu/abs/2016MNRAS.460.3472E} {460, 3472}

\bibitem[\protect\citeauthoryear{{Ercolano} \& {Pascucci}}{{Ercolano} \&
  {Pascucci}}{2017}]{2017RSOS....470114E}
{Ercolano} B.,  {Pascucci} I.,  2017, \mn@doi [Royal Society Open Science]
  {10.1098/rsos.170114}, \href
  {https://ui.adsabs.harvard.edu/abs/2017RSOS....470114E} {4, 170114}

\bibitem[\protect\citeauthoryear{{Fang} et~al.,}{{Fang}
  et~al.}{2018}]{2018ApJ...868...28F}
{Fang} M.,  et~al., 2018, \mn@doi [\apj] {10.3847/1538-4357/aae780}, \href
  {https://ui.adsabs.harvard.edu/abs/2018ApJ...868...28F} {868, 28}

\bibitem[\protect\citeauthoryear{{Fedele}, {van den Ancker}, {Henning},
  {Jayawardhana}  \& {Oliveira}}{{Fedele} et~al.}{2010}]{2010A&A...510A..72F}
{Fedele} D.,  {van den Ancker} M.~E.,  {Henning} T.,  {Jayawardhana} R.,
  {Oliveira} J.~M.,  2010, \mn@doi [\aap] {10.1051/0004-6361/200912810}, \href
  {https://ui.adsabs.harvard.edu/abs/2010A&A...510A..72F} {510, A72}

\bibitem[\protect\citeauthoryear{{Font}, {McCarthy}, {Johnstone}  \&
  {Ballantyne}}{{Font} et~al.}{2004}]{2004ApJ...607..890F}
{Font} A.~S.,  {McCarthy} I.~G.,  {Johnstone} D.,   {Ballantyne} D.~R.,  2004,
  \mn@doi [\apj] {10.1086/383518}, \href
  {https://ui.adsabs.harvard.edu/abs/2004ApJ...607..890F} {607, 890}

\bibitem[\protect\citeauthoryear{{Glassgold}, {Najita}  \& {Igea}}{{Glassgold}
  et~al.}{2007}]{2007ApJ...656..515G}
{Glassgold} A.~E.,  {Najita} J.~R.,   {Igea} J.,  2007, \mn@doi [\apj]
  {10.1086/510013}, \href
  {https://ui.adsabs.harvard.edu/abs/2007ApJ...656..515G} {656, 515}

\bibitem[\protect\citeauthoryear{{Gorti} \& {Hollenbach}}{{Gorti} \&
  {Hollenbach}}{2009}]{2009ApJ...690.1539G}
{Gorti} U.,  {Hollenbach} D.,  2009, \mn@doi [\apj]
  {10.1088/0004-637X/690/2/1539}, \href
  {https://ui.adsabs.harvard.edu/abs/2009ApJ...690.1539G} {690, 1539}

\bibitem[\protect\citeauthoryear{{Haisch}, {Lada}  \& {Lada}}{{Haisch}
  et~al.}{2001}]{2001ApJ...553L.153H}
{Haisch} Jr. K.~E.,  {Lada} E.~A.,   {Lada} C.~J.,  2001, \mn@doi [\apjl]
  {10.1086/320685}, \href
  {https://ui.adsabs.harvard.edu/abs/2001ApJ...553L.153H} {553, L153}

\bibitem[\protect\citeauthoryear{{Hartigan}, {Edwards}  \&
  {Ghandour}}{{Hartigan} et~al.}{1995}]{1995ApJ...452..736H}
{Hartigan} P.,  {Edwards} S.,   {Ghandour} L.,  1995, \mn@doi [\apj]
  {10.1086/176344}, \href
  {https://ui.adsabs.harvard.edu/abs/1995ApJ...452..736H} {452, 736}

\bibitem[\protect\citeauthoryear{{Hartmann}, {Calvet}, {Gullbring}  \&
  {D'Alessio}}{{Hartmann} et~al.}{1998}]{1998ApJ...495..385H}
{Hartmann} L.,  {Calvet} N.,  {Gullbring} E.,   {D'Alessio} P.,  1998, \mn@doi
  [\apj] {10.1086/305277}, \href
  {https://ui.adsabs.harvard.edu/abs/1998ApJ...495..385H} {495, 385}

\bibitem[\protect\citeauthoryear{{Haworth} \& {Clarke}}{{Haworth} \&
  {Clarke}}{2019}]{2019MNRAS.485.3895H}
{Haworth} T.~J.,  {Clarke} C.~J.,  2019, \mn@doi [\mnras]
  {10.1093/mnras/stz706}, \href
  {https://ui.adsabs.harvard.edu/abs/2019MNRAS.485.3895H} {485, 3895}

\bibitem[\protect\citeauthoryear{{Hollenbach} \& {Gorti}}{{Hollenbach} \&
  {Gorti}}{2009}]{2009ApJ...703.1203H}
{Hollenbach} D.,  {Gorti} U.,  2009, \mn@doi [\apj]
  {10.1088/0004-637X/703/2/1203}, \href
  {https://ui.adsabs.harvard.edu/abs/2009ApJ...703.1203H} {703, 1203}

\bibitem[\protect\citeauthoryear{{Hollenbach}, {Johnstone}, {Lizano}  \&
  {Shu}}{{Hollenbach} et~al.}{1994}]{1994ApJ...428..654H}
{Hollenbach} D.,  {Johnstone} D.,  {Lizano} S.,   {Shu} F.,  1994, \mn@doi
  [\apj] {10.1086/174276}, \href
  {https://ui.adsabs.harvard.edu/abs/1994ApJ...428..654H} {428, 654}

\bibitem[\protect\citeauthoryear{{Ingleby} et~al.,}{{Ingleby}
  et~al.}{2009}]{2009ApJ...703L.137I}
{Ingleby} L.,  et~al., 2009, \mn@doi [\apjl] {10.1088/0004-637X/703/2/L137},
  \href {https://ui.adsabs.harvard.edu/abs/2009ApJ...703L.137I} {703, L137}

\bibitem[\protect\citeauthoryear{{Koepferl}, {Ercolano}, {Dale}, {Teixeira},
  {Ratzka}  \& {Spezzi}}{{Koepferl} et~al.}{2013}]{2013MNRAS.428.3327K}
{Koepferl} C.~M.,  {Ercolano} B.,  {Dale} J.,  {Teixeira} P.~S.,  {Ratzka} T.,
   {Spezzi} L.,  2013, \mn@doi [\mnras] {10.1093/mnras/sts276}, \href
  {https://ui.adsabs.harvard.edu/abs/2013MNRAS.428.3327K} {428, 3327}

\bibitem[\protect\citeauthoryear{{Liffman}}{{Liffman}}{2003}]{2003PASA...20..337L}
{Liffman} K.,  2003, \mn@doi [\pasa] {10.1071/AS03019}, \href
  {https://ui.adsabs.harvard.edu/abs/2003PASA...20..337L} {20, 337}

\bibitem[\protect\citeauthoryear{{Mathews}, {Williams}, {M{\'e}nard},
  {Phillips}, {Duch{\^e}ne}  \& {Pinte}}{{Mathews}
  et~al.}{2012}]{2012ApJ...745...23M}
{Mathews} G.~S.,  {Williams} J.~P.,  {M{\'e}nard} F.,  {Phillips} N.,
  {Duch{\^e}ne} G.,   {Pinte} C.,  2012, \mn@doi [\apj]
  {10.1088/0004-637X/745/1/23}, \href
  {https://ui.adsabs.harvard.edu/abs/2012ApJ...745...23M} {745, 23}

\bibitem[\protect\citeauthoryear{{Natta}, {Testi}, {Alcal{\'a}}, {Rigliaco},
  {Covino}, {Stelzer}  \& {D'Elia}}{{Natta} et~al.}{2014}]{2014A&A...569A...5N}
{Natta} A.,  {Testi} L.,  {Alcal{\'a}} J.~M.,  {Rigliaco} E.,  {Covino} E.,
  {Stelzer} B.,   {D'Elia} V.,  2014, \mn@doi [\aap]
  {10.1051/0004-6361/201424136}, \href
  {https://ui.adsabs.harvard.edu/abs/2014A&A...569A...5N} {569, A5}

\bibitem[\protect\citeauthoryear{{Owen}, {Ercolano}, {Clarke}  \& {Alexand
  er}}{{Owen} et~al.}{2010}]{2010MNRAS.401.1415O}
{Owen} J.~E.,  {Ercolano} B.,  {Clarke} C.~J.,   {Alexand er} R.~D.,  2010,
  \mn@doi [\mnras] {10.1111/j.1365-2966.2009.15771.x}, \href
  {https://ui.adsabs.harvard.edu/abs/2010MNRAS.401.1415O} {401, 1415}

\bibitem[\protect\citeauthoryear{{Pascucci} \& {Sterzik}}{{Pascucci} \&
  {Sterzik}}{2009}]{2009ApJ...702..724P}
{Pascucci} I.,  {Sterzik} M.,  2009, \mn@doi [\apj]
  {10.1088/0004-637X/702/1/724}, \href
  {https://ui.adsabs.harvard.edu/abs/2009ApJ...702..724P} {702, 724}

\bibitem[\protect\citeauthoryear{{Pascucci} et~al.,}{{Pascucci}
  et~al.}{2006}]{2006ApJ...651.1177P}
{Pascucci} I.,  et~al., 2006, \mn@doi [\apj] {10.1086/507761}, \href
  {https://ui.adsabs.harvard.edu/abs/2006ApJ...651.1177P} {651, 1177}

\bibitem[\protect\citeauthoryear{{Pascucci} et~al.,}{{Pascucci}
  et~al.}{2007}]{2007ApJ...663..383P}
{Pascucci} I.,  et~al., 2007, \mn@doi [\apj] {10.1086/518535}, \href
  {https://ui.adsabs.harvard.edu/abs/2007ApJ...663..383P} {663, 383}

\bibitem[\protect\citeauthoryear{{Pascucci} et~al.,}{{Pascucci}
  et~al.}{2011}]{2011ApJ...736...13P}
{Pascucci} I.,  et~al., 2011, \mn@doi [\apj] {10.1088/0004-637X/736/1/13},
  \href {https://ui.adsabs.harvard.edu/abs/2011ApJ...736...13P} {736, 13}

\bibitem[\protect\citeauthoryear{{Picogna}, {Ercolano}, {Owen}  \&
  {Weber}}{{Picogna} et~al.}{2019}]{2019MNRAS.487..691P}
{Picogna} G.,  {Ercolano} B.,  {Owen} J.~E.,   {Weber} M.~L.,  2019, \mn@doi
  [\mnras] {10.1093/mnras/stz1166}, \href
  {https://ui.adsabs.harvard.edu/abs/2019MNRAS.487..691P} {487, 691}

\bibitem[\protect\citeauthoryear{{Rigliaco}, {Pascucci}, {Gorti}, {Edwards}  \&
  {Hollenbach}}{{Rigliaco} et~al.}{2013}]{2013ApJ...772...60R}
{Rigliaco} E.,  {Pascucci} I.,  {Gorti} U.,  {Edwards} S.,   {Hollenbach} D.,
  2013, \mn@doi [\apj] {10.1088/0004-637X/772/1/60}, \href
  {https://ui.adsabs.harvard.edu/abs/2013ApJ...772...60R} {772, 60}

\bibitem[\protect\citeauthoryear{{Sacco} et~al.,}{{Sacco}
  et~al.}{2012}]{2012ApJ...747..142S}
{Sacco} G.~G.,  et~al., 2012, \mn@doi [\apj] {10.1088/0004-637X/747/2/142},
  \href {https://ui.adsabs.harvard.edu/abs/2012ApJ...747..142S} {747, 142}

\bibitem[\protect\citeauthoryear{{Simon} \& {Prato}}{{Simon} \&
  {Prato}}{1995}]{1995ApJ...450..824S}
{Simon} M.,  {Prato} L.,  1995, \mn@doi [\apj] {10.1086/176187}, \href
  {https://ui.adsabs.harvard.edu/abs/1995ApJ...450..824S} {450, 824}

\bibitem[\protect\citeauthoryear{{Simon}, {Pascucci}, {Edwards}, {Feng},
  {Gorti}, {Hollenbach}, {Rigliaco}  \& {Keane}}{{Simon}
  et~al.}{2016}]{2016ApJ...831..169S}
{Simon} M.~N.,  {Pascucci} I.,  {Edwards} S.,  {Feng} W.,  {Gorti} U.,
  {Hollenbach} D.,  {Rigliaco} E.,   {Keane} J.~T.,  2016, \mn@doi [\apj]
  {10.3847/0004-637X/831/2/169}, \href
  {https://ui.adsabs.harvard.edu/abs/2016ApJ...831..169S} {831, 169}

\bibitem[\protect\citeauthoryear{{Wang} \& {Goodman}}{{Wang} \&
  {Goodman}}{2017}]{2017ApJ...847...11W}
{Wang} L.,  {Goodman} J.,  2017, \mn@doi [\apj] {10.3847/1538-4357/aa8726},
  \href {https://ui.adsabs.harvard.edu/abs/2017ApJ...847...11W} {847, 11}

\makeatother
\end{thebibliography}




\appendix




\bsp	
\label{lastpage}
\end{document}